\documentclass[12pt,onecolumn,oneside,a4paper,peerreview]{IEEEtran}
\usepackage{cite}
\usepackage{graphicx}
\usepackage{amsmath}
\usepackage{times}
\usepackage{latexsym}
\usepackage{graphicx}
\usepackage{bm}
\usepackage{amssymb}
\usepackage[center]{caption2}
\usepackage{stfloats}
\usepackage{cases}
\usepackage{subfigure}
\usepackage{array}
\usepackage{setspace}
\usepackage{fancyhdr}
\usepackage{cite}
\usepackage{color}

\newtheorem{theorem}{Theorem}
\newtheorem{lemma}{Lemma}

\newtheorem{proposition}{Proposition}
\newtheorem{definition}{Definition}
\newtheorem{remark}{Remark}
\def\proof{\noindent\hspace{2em}{\itshape Proof: }}
\def\endproof{\hspace*{\fill}~$\square$\par\endtrivlist\unskip}

\newcaptionstyle{mystyle2}{%
\captionlabel $.$ \, \doublespacing \captiontext \par}
\captionstyle{mystyle2}

\begin{document}
\title{Wireless Powered Dual-Hop Multi-Antenna Relaying Systems: Impact of CSI and Antenna Correlation}

\author{Han Liang,~\IEEEmembership{Student Member,~IEEE,} Caijun Zhong,~\IEEEmembership{Senior Member,~IEEE,} Xiaoming Chen,~\IEEEmembership{Senior Member,~IEEE,} Himal A. Suraweera,~\IEEEmembership{Senior Member,~IEEE}, and Zhaoyang Zhang,~\IEEEmembership{Member,~IEEE}
      \thanks{H. Liang, C. Zhong and Z. Zhang are with the Institute of Information and Communication Engineering, Zhejiang University, China (email: caijunzhong@zju.edu.cn).}
\thanks{X. Chen is with the College of Electronic and Information Engineering, Nanjing University of Aeronautics and Astronautics, Nanjing, China (e-mail: chenxiaoming@nuaa.edu.cn).}
\thanks{H. A. Suraweera is with the Department of Electrical \& Electronic Engineering, University of Peradeniya, Peradeniya 20400, Sri Lanka (e-mail: himal@ee.pdn.ac.lk).}
}
\maketitle

\begin{abstract}
This paper investigates the impact of the channel state information (CSI) and antenna correlation at the multi-antenna relay on the performance of wireless powered dual-hop amplify-and-forward relaying systems. Depending on the available CSI at the relay, two different scenarios are considered, namely, instantaneous CSI and statistical CSI where the relay has access only to the antenna correlation matrix. Adopting the power-splitting architecture, we present a detailed performance study for both cases. Closed-form analytical expressions are derived for the outage probability and ergodic capacity. In addition, simple high signal-to-noise ratio (SNR) outage approximations are obtained. Our results show that, antenna correlation itself does not affect the achievable diversity order, the availability of CSI at the relay determines the achievable diversity order. Full diversity order can be achieved with instantaneous CSI, while only a diversity order of one can be achieved with statistical CSI. In addition, the transmit antenna correlation and receive antenna correlation exhibit different impact on the ergodic capacity. Moreover, the impact of antenna correlation on the ergodic capacity also depends heavily on the available CSI and operating SNR.
\end{abstract}


\newpage
\section{Introduction}\label{section:introduction}
How to prolong the operation lifetime of energy constrained wireless devices has become a critical issue, especially with the explosive growth of sensor nodes due to the fast development of internet of things. Responding to this challenge, empowering wireless devices with energy harvesting capabilities has been proposed as a promising solution \cite{C.K.Ho,M.Pinuela,S.Luo}, where the wireless devices can be designed to scavenge energy from natural resources such as solar and wind. However, the energy harvested from such sources is random and intermittent, and highly depends on uncontrollable factors such as weather, making it undesirable for communication systems with stringent quality-of-service (QoS) requirements \cite{K.Ioannis}.

As a practically viable solution to address this challenge, a new paradigm has emerged where the wireless devices harvest energy from the radio-frequency (RF) signals \cite{Dusit}. Since the RF signals can be fully controlled, it is more reliable and efficient. The combination of wireless power and information transfer has resulted in a new topic, generally referred to as \emph{simultaneous wireless information and power transfer} (SWIPT). The idea of SWIPT was initially proposed by Varshney in \cite{LavR}, where the author studied the fundamental tradeoff between the information capacity and the harvested energy under the ideal assumption that the circuit can decode the information and harvest energy at the same time. Later in \cite{R.Zhang}, two practical receiver architectures were proposed, namely time-switching and power-splitting. A more sophisticated dynamic power splitting scheme was proposed in \cite{L.Liu}. The performance of wirelessly powered multiple antenna systems with energy beamforming was studied in \cite{caijun,W.Huang}, the issue of imperfect channel state information (CSI) was addressed in \cite{Z.Xiang}, and a training based SWIPT system was studied in \cite{X.Zhou0}. {In addition, the performance of SWIPT in cellular system was considered in \cite{Derrick,Q.Wu} and the hybrid conventional battery and wireless energy transfer systems were investigated in \cite{Q.Wu0,Q.Wu01}.}


{SWIPT also finds important applications in cooperative relaying system \cite{A.Nasir,Z.Ding0,H.Chen} which is a fundamental building block for many important deployment scenarios in 5G systems such as Internet of things \cite{Q.Wu2}.} The authors in \cite{A.Nasir} investigated the throughput performance of amplify-and-forward (AF) half-duplex (HF) relaying network for both time-switching and power-splitting receiver architectures in Rayleigh fading channels, and \cite{Y.chen} extended the analysis to the more general Nakagami-m fading channel. The authors in \cite{Z.Ding1} considered the decode-and-forward relaying system and studied the power allocation strategies for multiple source-destination pairs. Later in \cite{I.Krikidis}, the authors proposed a low-complexity antenna switching protocol for the implementation of SWIPT, while in \cite{D.S.Mich}, the authors studied the performance of relay selection in SWIPT systems. The performance of energy harvesting cooperative networks with randomly distributed users/relays was studied in \cite{Z.Ding2,I.Krikidis}. In \cite{C.Zhong}, full-duplex (FD) relaying was introduced into SWIPT systems, while \cite{Trung,D.Li} considered the extension of two-way relaying. More recently, the impact of multiple antennas on the performance of dual-hop AF SWIPT relaying system was studied in \cite{G.Zhu}, where it was shown that increasing the number of relay antennas can substantially improve the system performance. However, the conclusion of \cite{G.Zhu} is based on the assumption that the multiple antennas are uncorrelated and the instantaneous CSI is available at the relay. In practice, due to insufficient antenna spacing and lack of local scatters, the channels tend to exhibit spatial correlation. And in the cases that channels are rapidly time-varying, in order to reduce the complexity of channel estimation, the relay may obtain only statistical CSI rather than instantaneous CSI. While the effect of statistical CSI has been extensively investigated in conventional multiple antenna systems\cite{X.Li,A.Zappone}, it has not been investigated in the dual-hop multi-antenna SWIPT systems.\footnote{In the conference paper \cite{liang.conf}, we have investigated the impact of CSI and antenna correlation on the outage probability of the dual-hop multi-antenna SWIPT systems. However, the detailed proof of the key results is not included due to space limitation, which will be presented here.} Therefore, it is of great interest to investigate the achievable performance of SWIPT relaying systems in realistic scenarios taking into consideration the impact of antenna correlation and statistical CSI.

Motivated by this, we consider a source-relay-destination dual-hop system where the source and destination are equipped with a single antenna while the relay powered via RF energy harvesting is equipped with multiple antennas. Unlike \cite{G.Zhu}, we assume that the antennas are spatially correlated which is practical in the situation that the space among antennas is insufficient or the local scatters are inadequate. In addition, we consider two scenarios depending on the available CSI at the relay node, i.e., instantaneous CSI or statistical CSI where the relay only have access to the antenna correlation information.

The main contributions are summarized as follows:
\begin{itemize}
  \item For the instantaneous CSI scenario, an exact analytical expression in integral form is derived for the outage probability, and a closed-form lower bound for the outage probability is presented. Also, simple asymptotic approximations for the outage probability are presented in the high signal-to-noise ratio (SNR) regime, which shows that full diversity can be achieved.
 \item For the statistical CSI scenario, a suboptimal relay processing matrix is proposed. An exact analytical expression in integral form is derived for the outage probability, and simple high SNR approximations for the outage probability are presented, which reveals that only unit diversity order can be attained.
 \item For both the instantaneous and statistical CSI scenarios, closed-form upper bounds for the ergodic capacity are derived. In addition, the impact of transmit and receive antenna correlation on the ergodic capacity is characterized.
 \item The outcomes of the paper suggest that the amount of available CSI at the relay has a significant impact on the system performance. Moreover, the impact of antenna correlation on the system performance is also coupled with the availability of CSI at the relay. Specifically, the transmit antenna correlation is detrimental in the case of instantaneous CSI while it is beneficial with statistical CSI, while the impact of receive antenna correlation is undetermined which depends heavily on the availability of CSI and the operating SNR.
\end{itemize}


The rest of the paper is organized as follows: Section \ref{system model for perfect CSI} introduces the system model, Section \ref{section2} focuses on the instantaneous CSI scenario, while Section \ref{section imperfect csi model} deals with the statistical CSI scenario. Numerical results are presented in Section \ref{numerical}. Finally, Section \ref{section conclusion} concludes the paper and summarizes the main findings.

{\em Notation:} The upper bold case letters, lower bold case letters and lower case letters denote matrices, vectors and scalars respectively. For an arbitrary-size matrix $\bf{R}$, ${\bf{R}}^H$, ${\bf{R}}^{*}$ and ${\bf{R}}^{\frac{1}{2}}$ denote the conjugate transpose, conjugate and square-root of ${\bf{R}}$ respectively. $\left\|{{\bf{h}}}\right\|$ denotes the Frobenius norm of a complex vector ${\bf{h}}$. ${\bf{h}}_i$ denotes the $i$-th element of vector ${\bf{h}}$. ${F}_t(x)$ denotes the cumulative distribution function (CDF) of a random variable $t$. $\bar{F}_t(x)$ denotes $1-{F}_t(x)$. $E \left\{ \cdot \right\}$ denotes the statistical expectation. $\Gamma(x)$ is the gamma function\cite[Eq. (8.310.1)]{Tables}. ${\sf Ei}(x)$ is the exponential integral function \cite[Eq. (8.211.1)]{Tables}. $\psi(x)$ is the psi function\cite[Eq. (8.360)]{Tables}. $K_n\left(x\right)$ is the n-th order modified Bessel function of the second kind \cite[Eq. (8.407)]{Tables}. $G_{p,q}^{m,n}(\cdot)$ is the Meijer's G-function\cite[Eq. (9.301)]{Tables}.

\section{System Model}\label{system model for perfect CSI}
We consider a dual-hop energy harvesting relay system where a single-antenna source, S, communicates with a single-antenna destination, D, with the aid of a multi-antenna relay, R, as illustrated in Fig. 1(a). It is assumed that the relay has no external power supply, and entirely relies on the energy harvested from the source signal. \footnote{In this paper, we have assumed that the processing power required by the transmit/receive circuitry at the relay is negligible as compared to the relay transmission power. This assumption is justifiable when the transmission distances are relatively large such that the transmission energy becomes the dominant source of energy consumption \cite{X.Zhou2}. As a matter of fact, similar assumption has been widely adopted in SWIPT relaying literature with both single-antenna relay \cite{A.Nasir} and multi-antenna relay \cite{Y.Huang}. In addition, as in \cite{A.Nasir,Amhed}, we assume that the relay is entirely powered via RF energy harvesting. By implementing multiple antennas at the relay, the energy harvesting efficiency can be significantly improved \cite{U.Olgun,K.Ioannis}, hence more energy can be harvested to support its power requirements. Therefore, we believe that adopting multi-antenna relay is practically viable.} We assume that the direct link between the source and destination does not exist due to obstacles or severe shadowing as in \cite{A.Nasir,C.Zhong,G.Zhu}. {{ It is assumed that CSI is available at the destination node, while no CSI is available at the source node as in \cite{G.Zhu,R.Zhang2}. Also, two different types of CSI assumptions are considered at the relay node, namely, instantaneous CSI and statistical CSI. In practice, CSI can be obtained by pilot-assisted channel training or with the help of a suitable feedback mechanism.}}

We focus on the power splitting receiver architecture as in \cite{A.Nasir}. Hence, an entire transmission block is divided into two time slots with duration of \(\frac{T}{2}\) each. During the first time slot, the relay listens to the source transmission, and splits the received signal into a power stream for energy harvesting and an information stream for information forwarding according to a power splitting ratio $\theta$. During the second time slot, the relay forwards the processed information to the destination using the harvested power.

Let the $N\times 1$ vector ${\bf{h}}_1$ and $1\times N$ vector ${\bf{h}}_2$ represent the channel coefficients of the source-relay and relay-destination links, respectively. We assume that channels are correlated due to insufficient antenna spacing, as such, ${\bf{h}}_1={\bf{R}}_r^\frac{1}{2}{\bf{h}}_{w1}$ and ${\bf{h}}_2={\bf{h}}_{w2}{\bf{R}}_t^\frac{1}{2}$, where $\textbf{R}_r$ and $\textbf{R}_t$ denote the receive and transmit correlation matrices, respectively. ${\bf{h}}_{w1}$ and ${\bf{h}}_{w2}$ are mutually independent zero-mean complex circular symmetric Gaussian random vectors with unit variance. Without loss of generality, the correlation matrices are assumed to be full rank.\footnote{To simplify the analytical derivation, we only provide the results for the full-rank case. However, the results for the arbitrary rank case can be obtained in a similar fashion, albeit with more involved mathematical manipulations.}

\begin{figure}[ht]
  \centering
  \includegraphics[scale=0.4]{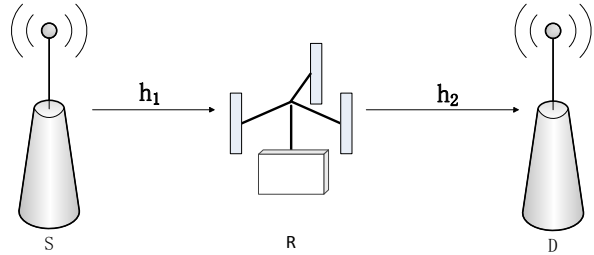}

  \caption{The considered dual-hop multiple antenna energy harvesting system.}
  \label{fig:1}
\end{figure}

As a result, the total harvested energy at the relay during the first time slot can be computed as
\begin{align}
{E_h} = {\frac{\eta\theta P_s}{d_1^\tau}}{\left\|{\bf{h}}_1\right\|^2}{\frac{T}{2}},
\end{align}
where $0<\eta\leqslant1$ denotes the energy conversion efficiency, $P_s$ denotes the source transmit power, $d_1$ is the distance between the source and relay, $\tau$ denotes the pass loss exponent.

Also, the signal received by the information receiver of the relay can be written as
\begin{align}\label{SM:1}
{{\bf{y}}_r}=\sqrt{{\frac{(1-\theta)P_s}{d_1^\tau}}}{\bf{h}}_1 x+{\bf{n}}_r,
\end{align}
where $x$ denotes the source symbol with unit power, and ${\bf{n}}_r$ denotes the additive white Gaussian noise (AWGN) satisfying ${E\{{\bf{n}}_r{\bf{n}}_r^H\}} = N_0{\textbf{\emph{I}}}$ with ${\textbf{\emph{I}}}$ being the identity matrix.

During the second time slot, the relay employs the AF protocol to forward the transformed signal to the destination using the harvested energy. Depending on the availability of CSI at the relay, we consider two separate scenarios, namely, instantaneous CSI and statistical CSI.
\subsection{Instantaneous CSI}
In this case, it is assumed that the relay knows both ${\bf h}_1$ and ${\bf h}_2$. With the AF protocol, the relay simply scales the received signal by a transformation matrix ${\bf{G}}$. To meet the power constraint at the relay, it is required that ${E\left\{\left\|{\bf{G}}{\bf{y}}_r\right\|^2\right\}}={P_r}$, where $P_r=\frac{2 E_h}{T}$. According to \cite{G.Zhu.two}, the optimal transformation matrix maximizing the end-to-end SNR can be expressed as
\begin{align}
{\bf{G}} = \omega\frac{{\bf{h}}_2^H{\bf{h}}_1^H}{\left\|{\bf{h}}_2\right\|\left\|{\bf{h}}_1\right\|},
\end{align}
where $\omega$ is the power constraint factor and is given by
\begin{align}\label{omegasqure}
\omega^2 = \frac{\frac{\eta\theta P_s}{d_1^\tau}\left\|{\bf{h}}_1\right\|^2}{\frac{(1-\theta)P_s}{d_1^\tau}\left\|{\bf{h}}_1\right\|^2+N_0}.
\end{align}
Therefore, the received signal at the destination can be expressed as
\begin{align}
y_d = \sqrt{\frac{1}{d_2^\tau}}{{\bf{h}}_2}{\bf{G}}{{\bf{y}}_r}+{n_d},
\end{align}
where $d_2$ denotes the distance between the relay and the destination node, $n_d$ denotes the AWGN at the destination with $E\{n_r n_r^*\}=N_0$.

Define $\rho\triangleq\frac{P_s}{N_0}$, then the end-to-end signal-to-noise ratio (SNR) can be written as
\begin{align}\label{SNR 1}
\gamma^{\sf}_{I}=\frac{\frac{\eta\theta(1-\theta)\rho^2}{d_1^{2\tau}d_2^\tau}\left\|{\bf{h}}_2\right\|^2\left\|{\bf{h}}_1\right\|^4}{\frac{\eta\theta\rho}{d_1^\tau d_2^\tau}\left\|{\bf{h}}_2\right\|^2\left\|{\bf{h}}_1\right\|^2+\frac{(1-\theta)\rho}{d_1^\tau}\left\|{\bf{h}}_1\right\|^2+1}.
\end{align}

\subsection{Statistical CSI}
In this case, it is assumed that the relay only knows the channel correlation matrices $\textbf{R}_t$ and $\textbf{R}_r$. With only channel correlation information available, analytical characterization of the optimal relay processing matrix ${\bf{G}}$ appears to be difficult. Hence, we adopt a heuristic Rank-1 processing matrix ${\bf{G}}$ as motivated by the instantaneous CSI scenario,\footnote{{While the rank-1 processing matrix is a heuristic choice, we believe that the achievable system performance with rank-1 constraint should come close to that of the optimal solution. The main reason is that this paper assumes single antenna at the both the source and destination, hence, the single stream transmission in both the first hop and second hop is a reasonable choice, as demonstrated in prior works \cite{P.Dharmawansa,B.Khoshnevis}.}} i.e.,
\begin{align}
{\bf{G}}_{opt}=\omega{\bf{w}}_t{\bf{w}}_r,
\end{align}
where $\left\|{\bf{w}}_t\right\|^2=1$, $\left\|{\bf{w}}_r\right\|^2=1$ and $\omega$ is the power constraint.

With the Rank-1 constraint, we have the following important observation:
\begin{theorem}\label{optimal scheme}
With the Rank-1 assumption, the optimal ${\bf{w}}_t$ and ${\bf{w}}_r$ maximizing the ergodic capacity turn out to be the eigenvectors associated with the maximum eigenvalues of the correlation matrix ${\bf{R}}_t$ and ${\bf{R}}_r$, respectively.
\end{theorem}

\proof
See Appendix \ref{appendix:optimal scheme}.
\endproof

Now, consider the following eigenvalue decomposition of the correlation matrices ${\bf{R}}_r={\bf{U}}_r\Sigma_r{\bf{U}}_r^{H}$ and ${\bf{R}}_t={\bf{U}}_t\Sigma_t{\bf{U}}_t^{H}$, then, we have
\begin{align}
\left|{\bf{w}}_r{\bf{h}}_1\right|^2=\lambda_1\left|\tilde{h}_{w11}\right|^2 \quad \mbox{and}\quad \left|{\bf{h}}_2{\bf{w}}_t\right|^2=\sigma_1\left|\tilde{h}_{w21}\right|^2,
\end{align}
where $\tilde{h}_{w1i}$ and $\tilde{h}_{w2m}$ denote the $i$-th and $m$-th element of ${\bf{U}}_r{\bf{h}}_{w1}$ and ${\bf{h}}_{w2}{\bf{U}}_t$, respectively. And $\lambda_i$ and $\sigma_m$ are the $i$-th and $m$-th largest eigenvalues of ${\bf{R}}_r$ and ${\bf{R}}_t$, respectively. Then, the end-to-end SNR can be derived as
\begin{align}\label{inscsi snr}
\gamma_{S}=\frac{\frac{\eta\theta(1-\theta)\rho^2}{d_1^{2\tau}d_2^\tau}\left\|{\bf{h}}_1\right\|^2\lambda_1\left|\tilde{h}_{w11}\right|^2\sigma_1\left|\tilde{h}_{w21}\right|^2}{\frac{\eta\theta\rho}{d_1^\tau d_2^\tau}\left\|{\bf{h}}_1\right\|^2\sigma_1\left|\tilde{h}_{w21}\right|^2+\frac{(1-\theta)\rho}{d_1^\tau}\lambda_1\left|\tilde{h}_{w11}\right|^2+1}.
\end{align}


\section{Instantaneous CSI Scenario}\label{section2}
In this section, we focus on the scenario with instantaneous CSI, and present a detailed analysis on the system performance in terms of key performance measures, namely, the outage probability and the ergodic capacity.

\subsection{Outage Probability}
Mathematically, the outage probability is defined as the probability of the instantaneous SNR falls below a pre-defined threshold $\gamma_{\sf th}$, i.e.,
\begin{align}\label{outage definition}
P_{\sf out}=P\left\{\gamma_{\sf I} < \gamma_{\sf th}\right\}.
\end{align}

\begin{proposition}
The exact outage probability can be expressed as
\begin{align}\label{outage}
P_{I}=1-\int_{d/c}^{\infty}f_{\left\|{\bf{h}}_1\right\|^2}(x)\bar{F}_{\left\|{\bf{h}}_2\right\|^2}\left(\frac{ax+b}{cx^2-dx}\right)dx,
\end{align}
where $a = \frac{(1-\theta)\rho\gamma_{\sf th}}{d_1^\tau}$, $b = \gamma_{\sf th}$, $c = \frac{\eta\theta(1-\theta)\rho^2}{d_1^{2\tau}d_2^\tau}$, $d = \frac{\eta\theta\rho\gamma_{\sf th}}{d_1^\tau d_2^\tau}$.
\end{proposition}
\proof The desired result can be obtained by following the similar lines as in the proof \cite{A.Nasir}.\endproof

Although the integral in (\ref{outage}) does admit a closed-form solution, it can be efficiently evaluated through numerical integration. Alternatively, we can use the following tight lower bound of the outage probability.

\begin{theorem}\label{theorem:1}
The outage probability of the system can be lower bounded as
\begin{align}\label{lower bound h1 h2}
P_{I}^{low} = 1-2\sum_{i=1}^{N}\sum_{m=1}^Ne^{-\frac{\gamma_{\sf th}d_1^\tau}{(1-\theta)\rho\lambda_i}}
\prod_{\begin{subarray}{l}j=1\\j\ne i\end{subarray}}^N\frac{\lambda_i^{N-1}}{\lambda_i-\lambda_j}
\prod_{\begin{subarray}{l}n=1\\n\ne m\end{subarray}}^N{\frac{\sigma_m^{N-1}}{\sigma_m-\sigma_n}}
2\sqrt{\frac{\gamma_{\sf th}d_1^\tau d_2^\tau}{\eta\theta\rho\lambda_i\sigma_m}}K_1\left(2\sqrt{\frac{\gamma_{\sf th}d_1^\tau d_2^\tau}{\eta\theta\rho\lambda_i\sigma_m}}\right).
\end{align}

\proof  See Appendix \ref{appendix:theorem:1}. \endproof
\end{theorem}

While Theorem \ref{theorem:1} gives an efficient method for evaluating a tight lower bound of the outage probability of the system, the expression in (\ref{lower bound h1 h2}) provides little insights. Motivated by this, we now look into the high SNR regime, and derive a simple approximation for the outage probability, which enables the characterization of the achievable diversity order.


\begin{theorem}\label{theorem:2}
In high SNR regime, the outage probability of the system can be approximated as
\begin{multline}\label{high snr h1h2}
\hspace{-0.3cm}P_{I}^{\rho\rightarrow\infty}\approx \sum_{i=1}^{N}\sum_{m=1}^N\prod_{\begin{subarray}{l}j=1\\ j\ne i\end{subarray}}^N\frac{\lambda_i^{-1}}{\lambda_i-\lambda_j}\prod_{\begin{subarray}{l}n=1\\n\ne m\end{subarray}}^N{\frac{\sigma_m^{-1}}{\sigma_m-\sigma_n}}\frac{1}{N!}\times\\
\hspace{-0.3cm}\left[\frac{1}{(N-1)!}\left(\frac{d_2^\tau}{\eta\theta}\right)^N\left(\sum_{l=1}^{N-1}\frac{1}{l}-C+\ln{\frac{(1-\theta)\rho\lambda_i}{\gamma_{\sf th}d_1^\tau}}\right)-\left(\frac{-\sigma_m}{1-\theta}\right)^N\right]\left(\frac{\gamma_{\sf th}d_1^\tau}{\rho}\right)^N.
\end{multline}

\proof See Appendix \ref{appendix:theorem:2}.\endproof
\end{theorem}

Theorem \ref{theorem:2} indicates that the system achieves the full diversity of $N$, which implies that antenna correlation does not reduce the diversity order, instead, it degrades the system performance by affecting the coding gain. Moreover, it is observed that $P_{\sf out}$ decays as $\rho^{-N}\ln{\rho}$ rather than $\rho^{-N}$, which suggests that the outage probability decays much slower than the conventional case where the relay has a constant power source, as previously reported in \cite{G.Zhu}. The reason of this phenomenon is possibly that the transmit power at the relay is a random variable in SWIPT systems, which causes higher outage probability compared to the conventional cases.

\subsection{Ergodic Capacity}\label{section3}
We now study the ergodic capacity of the system. To start with, we first re-write the end-to-end SNR in (\ref{SNR 1}) as
\begin{align}
\gamma_I = \frac{\gamma_1 \gamma_2}{\gamma_1 + \gamma_2 +1},
\end{align}
where $\gamma_1 = \frac{(1-\theta)\rho}{d_1^\tau}\left\|{\bf{h}}_1\right\|^2$  and $\gamma_2 = \frac{\eta\theta\rho}{d_1^\tau d_2^\tau}\left\|{\bf{h}}_2\right\|^2\left\|{\bf{h}}_1\right\|^2$. Hence, the ergodic capacity can be expressed as
\begin{align}\label{C}
C_{I} = \frac{1}{2}E\left[\log_2\left(1+\frac{\gamma_1 \gamma_2}{\gamma_1+\gamma_2+1}\right)\right].
\end{align}

Since the probability density function (PDF) of $\gamma_{I}$ is not available, exact characterization of the ergodic capacity is mathematically intractable. As such, we direct our efforts to seek tight capacity bounds. And we have the following key result:
\begin{theorem}\label{theorem:77}
The ergodic capacity of the system with the AF protocol can be upper bounded as
\begin{multline}\label{ergodic:3}
C_{I}^{up} = \frac{1}{2\ln{2}}\sum_{i=1}^N\prod_{\begin{subarray}{l}j=1\\j\ne i\end{subarray}}^N\frac{\lambda_i^{N-1}}{\lambda_i-\lambda_j}
\left[{\sf Ei}\left(-\frac{d_1^\tau}{(1-\theta)\rho\lambda_i}\right)\times\right.\\
\left(-e^{\frac{d_1^\tau}{(1-\theta)\rho\lambda_i}}\right)+\sum_{m=1}^N\prod_{\begin{subarray}{l}n=1\\n\ne m\end{subarray}}^N\frac{\sigma_m^{N-1}}{\sigma_m-\sigma_n}\left.G_{1 3}^{3 1}\left(\frac{d_1^\tau d_2^\tau}{\eta\theta\rho\lambda_i\sigma_m}\middle|_{0,1,0}^0\right)\right]-\frac{1}{2}\log_2{\left(1+e^{O_1}+e^{O_2}\right)},
\end{multline}
where
\begin{align}
\left\{
\begin{aligned}
O_1 =& \sum_{i=1}^N\prod_{\begin{subarray}{l}j=1\\ j\ne i\end{subarray}}^N\frac{\lambda_i^{N-1}}{\lambda_i-\lambda_j}\left(\psi(1)+\ln{\frac{(1-\theta)\rho\lambda_i}{d_1^\tau}}\right)\\
O_2 =& \sum_{i=1}^N\sum_{m=1}^{N-1}\prod_{\begin{subarray}{l}j=1\\j\ne i\end{subarray}}^N\frac{\lambda_i^{N-1}}{\lambda_i-\lambda_j}\prod_{\begin{subarray}{l}n=1\\n \ne m\end{subarray}}^N\frac{\sigma_m^{N-1}}{\sigma_m-\sigma_n}\left(2\psi(1)-\ln{\frac{d_1^\tau d_2^\tau}{\eta\theta\rho\lambda_i\sigma_m}}\right)
\end{aligned}
\right..
\end{align}

\proof
See Appendix \ref{appendix:theorem:77}.
 \endproof
\end{theorem}

Theorem \ref{theorem:77} presents a new upper bound for the ergodic capacity of the system, which is quite tight across the entire SNR range as shown in the section \ref{numerical}, hence, providing an efficient means to evaluate the capacity without resorting to time-consuming Monte Carlo simulations.

\subsection{Impact of Correlation on Ergodic Capacity}
We now study the impact of antenna correlation on the ergodic capacity. Before delving into the details, we first introduce the following definition given in \cite[Ch. 11]{A.W.Marshall}:
\begin{definition}\label{definition:majorize}
For two vectors ${\bf{a}}, {\bf{b}} \in \mathbb{R}^n$, we use ${\bf{a}}\succ{\bf{b}}$ to denote vector ${\bf{a}}$ majorize vector ${\bf{b}}$, if $\sum_{i=1}^k a_i \geqslant \sum_{i=1}^k b_i$, $k = 1,2\cdots,n-1$ and $\sum_{i=1}^n a_i = \sum_{i=1}^n b_i$, where $a_1\geqslant a_2 \geqslant \cdots \geqslant a_n$ and $b_1\geqslant b_2 \geqslant \cdots \geqslant b_n$ are the ordered elements of ${\bf{a}}$ and ${\bf{b}}$.
\end{definition}

The following proposition characterizes the impact of correlation on the ergodic capacity.
\begin{proposition}\label{proposition:impact of correlation on ergodic capacity}
The ergodic capacity of the system with instantaneous CSI $C_I$ is a Schur-concave function with respect to the eigenvalues of the transmit correlation matrix, i.e., let ${\bm{\sigma}}_1$ and ${\bm{\sigma}}_2$ be two vectors comprised of the eigenvalues of the transmit correlation matrix satisfying ${\bm{\sigma}}_1\succ{\bm{\sigma}}_2$, then we have


\begin{align}\label{eq:impact of co with inscsi}
C_I({\bm{\sigma}}_1)\leqslant C_I({\bm{\sigma}}_2).
\end{align}
{In addition, the high SNR ergodic capacity $C_I^h$ is a Schur-concave function, while the low SNR ergodic capacity $C_I^l$ is a Schur-convex function with respect to the eigenvalues of the receiver correlation matrix, i.e., let ${\bm{\lambda}}_1$ and ${\bm{\lambda}}_2$ be two vectors comprised of the eigenvalues of the receive correlation matrix satisfying ${\bm{\lambda}}_1\succ{\bm{\lambda}}_2$, then we have
\begin{align}\label{eq:impact of co with inscsi_rec}
C_I^h({\bm{\lambda}}_1)\leqslant C_I^h({\bm{\lambda}}_2), \mbox{ and }, C_I^l({\bm{\lambda}}_1)\geqslant C_I^l({\bm{\lambda}}_2).
\end{align}}
\end{proposition}
\proof
See Appendix \ref{appendix:impact of correlation on ergodic capacity}.
\endproof
Proposition \ref{proposition:impact of correlation on ergodic capacity} indicates that transmit correlation has a detrimental effect on the ergodic capacity, the stronger the correlation, the lower the ergodic capacity. Such an observation is in consistent with those reported in the MIMO literature. {Unfortunately, similar characterization on the impact of receive correlation is not available. In contrast, whether receive correlation is beneficial or detrimental depends on the operating SNR. At the high SNR regime, receive correlation leads to a reduction of the ergodic capacity, while at the low SNR regime, receive correlation contributes to an increase of the ergodic capacity.}


\section{Statistical CSI Scenario}\label{section imperfect csi model}
In this section, we present a detailed analysis on the system performance when the relay has only access to statistical CSI.

\subsection{Outage Probability}\label{section imperfect csi outage}
Due to the correlation between the random variables $\left\|{\bf{h}}_1\right\|^2$ and $\left|\tilde{h}_{w11}\right|^2$, the analysis becomes much more involved.
\begin{theorem}\label{theorem:imCSI AF outage}
With both transmit and receive correlation, the outage probability of the system with only statistical CSI at the relay is given by
\begin{align}\label{imCSI AF outage}
P_{S}=1-\int_{\frac{d}{c}}^{\infty}e^{-\left(\frac{at+b}{(ct-d)\sigma_1\lambda_1 t}+t\right)}dt-\sum_{i=2}^N\prod_{\begin{subarray}{l}j=2\\j \ne i\end{subarray}}^N\frac{\lambda_i^{N-2}}{\lambda_i-\lambda_j}\int_{\frac{d}{c}}^{\infty}e^{\left(\frac{\lambda_1}{\lambda_i}-1\right)t}\int_{\frac{d}{c}}^\frac{at+b}{(ct-d)\sigma_1\lambda_i t}e^{-\left(x+\frac{at+b}{(ct-d)\sigma_1\lambda_i x}\right)}dxdt.
\end{align}
\proof
See Appendix \ref{appendix:imCSI AF outage}.
\endproof
\end{theorem}

To the best of the authors' knowledge, (\ref{imCSI AF outage}) does not admit a closed-form expression. To gain further insights, we now derive a simple high SNR approximation for the outage probability, which enables the characterization of the achievable diversity order.
\begin{theorem}\label{theorem:imCSI AF high SNR}
In the high SNR region, the outage probability can be approximated as
\begin{align}\label{imCSI AF high SNR N>2}
P_{S}^{\rho\rightarrow \infty}=\left[\frac{\gamma_{\sf th}d_1^\tau}{(1-\theta)\lambda_1}-\sum_{i=1}^N\prod_{\begin{subarray}{l}j=1\\j\ne i\end{subarray}}^N\frac{\lambda_i^{N-2}}{\lambda_i-\lambda_j}\frac{d_1^\tau d_2^\tau \gamma_{\sf th}}{\eta\theta\sigma_1}\ln{\frac{d_1^\tau d_2^\tau \gamma_{\sf th}}{\eta\theta\sigma_1\lambda_i\rho}}\right]\frac{1}{\rho}.
\end{align}
\proof
See Appendix \ref{appendix:imCSI AF high SNR}.
\endproof
\end{theorem}

Theorem \ref{theorem:imCSI AF high SNR} suggests, with only the knowledge of correlation matrices at the relay, the system only achieves a diversity of one. Compared with the outage performance with instantaneous CSI in Section \ref{section2} where full diversity order is achieved, we can conclude the diversity order does not depend on the antenna correlation but is determined by the availability of CSI at the relay, and the correlation affects the system performance by affecting the coding gain. In addition, we observe that $P_{st}^{\rho\rightarrow \infty}$ in (\ref{imCSI AF high SNR N>2}) is a monotonic increasing function with respect to $\sigma_1$ which implies strong transmit correlation is beneficial in terms of the outage performance with only statistical CSI.

\subsection{Ergodic Capacity}\label{section imperfect csi ergodic}
In this section, we look into the ergodic capacity and have the following key result:
\begin{theorem}\label{theorem:imcsi ergodic}
With both transmit and receive correlation, the ergodic capacity of the system with only statistical CSI at the relay is upper bounded as
\begin{multline}
C_{S}^{up}=\frac{1}{2\ln{2}}\left[\sum_{i=1}^N\prod_{\begin{subarray}{l}j=1\\j\ne i\end{subarray}}^N\frac{\lambda_i^{N-1}}{\lambda_i-\lambda_j}G_{1 3}^{3 1}\left(\frac{d_1^\tau d_2^\tau}{\eta\theta\rho\sigma_1\lambda_i}\middle|_{0,1,0}^{0}\right)-e^{\frac{d_1^\tau}{(1-\theta)\rho\lambda_1}}{\sf Ei}\left(-\frac{d_1^\tau}{(1-\theta)\rho\lambda_1}\right)\right]\\-\frac{1}{2}\log_2\left(1+e^{O_1}+e^{O_2}\right),
\end{multline}
where
\begin{align}
\left\{\begin{aligned}
O_1&=\psi\left(1\right)+\ln{\frac{(1-\theta)\rho\lambda_1}{d_1^\tau}}\\
O_2&=\sum_{i=1}^N\prod_{\begin{subarray}{l}j=1\\j\ne i\end{subarray}}^N\frac{\lambda_i^{N-1}}{\lambda_i-\lambda_j}\left(2\psi(1)-\ln{\frac{d_1^\tau d_2^\tau \gamma_{\sf th}}{\eta\theta\rho\sigma_1\lambda_i}}\right)
\end{aligned}\right..
\end{align}
\proof
The result can be obtained by following similar lines as in proof of Theorem \ref{theorem:77}, along with some simple algebraic manipulations, hence is omitted.
\endproof
\end{theorem}
%
%
%
%

\subsection{Impact of Correlation on Ergodic Capacity}
We now study the impact of correlation on the ergodic capacity, and we have the following results:
\begin{proposition}\label{proposition impact of correlaion in statistical CSI}
The ergodic capacity of the system with statistical CSI $C_{S}$ is a Schur-convex function with respect to the eigenvalues of the transmit correlation matrix, i.e., let ${\bm{\sigma}}_1$ and ${\bm{\sigma}}_2$ be two vectors comprised of the eigenvalues of the transmit correlation matrix satisfying ${\bm{\sigma}}_1\succ{\bm{\sigma}}_2$, then we have
\begin{align}\label{eq:impact of co with stcsi}
C_S({\bm{\sigma}}_1)\geqslant C_S({\bm{\sigma}}_2).
\end{align}
For the special case $\lambda_{11}=\lambda_{21}$, $C_S$ is a Schur-concave function with respect to the eigenvalues of the receive correlation matrix, i.e., let ${\bm{\lambda}}_1$ and ${\bm{\lambda}}_2$ be two vectors comprised of the eigenvalues of the receive correlation matrix satisfying ${\bm{\lambda}}_1\succ{\bm{\lambda}}_2$ , then we have
\begin{align}\label{impact of correaltion st}
C_S({\bm{\lambda}}_1)\leqslant C_S({\bm{\lambda}}_2),
\end{align}
\proof
See Appendix \ref{appendix:proposition impact of correlaion in statistical CSI}.
\endproof

\end{proposition}

Please note, if $\lambda_{11}>\lambda_{21}$, then the impact of receive correlation is uncertain as shown in Appendix \ref{appendix:proposition impact of correlaion in statistical CSI}. In contrast to the instantaneous CSI scenario, Proposition \ref{proposition impact of correlaion in statistical CSI} indicates that transmit correlation is always beneficial with only statistical CSI. {The reason is that, with instantaneous CSI, a higher correlation reduces the array gain, which leads to a performance degradation. While with only statistical CSI, a higher correlation provides more information about the channel and enables power focusing, which contributes to performance enhancement.} {To this end, it is worth emphasizing that the above claims are established based on the rank-1 assumption. Hence, the impact of antenna correlation on the ergodic capacity with optimal relay processing matrix maybe different.}

\section{Numerical Results And Discussion}\label{numerical}
In this section, we present numerical results to demonstrate the correctness of the analytical expressions presented in Section \ref{section2} and Section \ref{section imperfect csi model}, and investigate the impact of key parameters on the system performance. Unless otherwise specified, we suppose ${\bf{h}}_1$ and ${\bf{h}}_2$ are exponentially-correlated\cite{S.Loyka} with different correlation factors $\lambda$ and $\sigma$ respectively, $\gamma_{\sf th}=0$ dB, $\eta = 0.8$, $\theta = 0.5$, $\tau = 2.5$ and $d_1 = d_2 = 3$ m.\footnote{{Please note, the choice of distances is simply for illustration purpose, and the developed analytical results are applicable for arbitrary system configurations.}}

To measure the impact of CSI on the system performance, we also include the case where no CSI is available at the relay as a benchmark scheme, where the relay processing matrix is given by ${\bf{G}}=w^2{\bf{I}}$, and the corresponding end-to-end SNR can be expressed as
\begin{align}\label{nocsi snr}
\gamma^{\sf}_{N}=\frac{\frac{\eta\theta(1-\theta)\rho^2}{d_1^{2\tau}d_2^\tau}\left\|{\bf{h}}_1\right\|^2\left|{\bf{h}}_2{\bf{h}}_1\right|^2}{\frac{\eta\theta\rho}{d_1^\tau d_2^\tau}\left\|{\bf{h}}_2\right\|^2\left\|{\bf{h}}_1\right\|^2+\frac{(1-\theta)\rho}{d_1^\tau}\left\|{\bf{h}}_1\right\|^2+1}.
\end{align}
Due to the challenge of characterizing the statistical properties of $\left|{\bf{h}}_2{\bf{h}}_1\right|^2$, the results for the no CSI case are obtained through Monte-Carlo simulations.

\begin{figure}[!ht]
  \centering
  \subfigure[Outage probability]{\label{fig:2a}\includegraphics[scale=0.5]{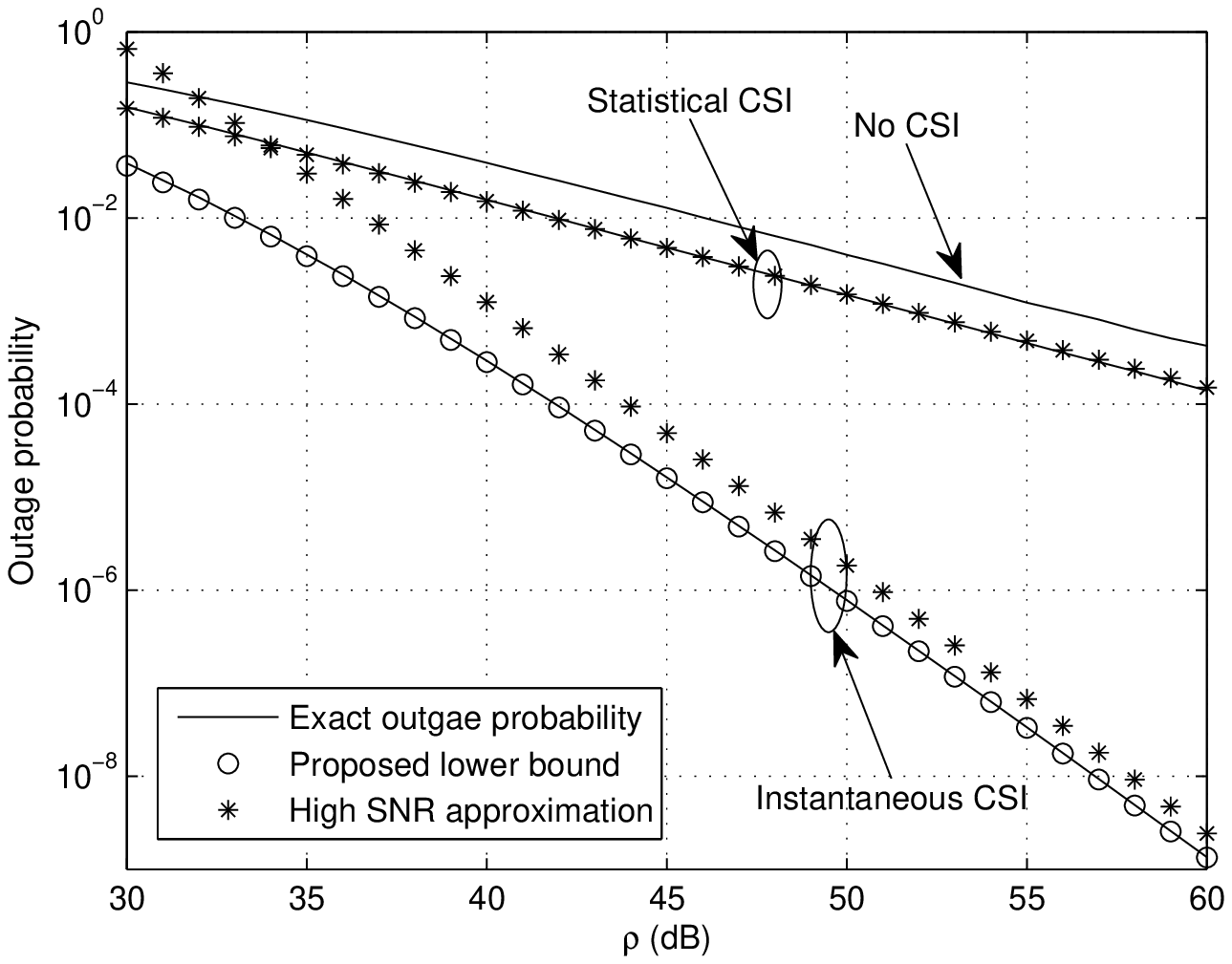}}
  \subfigure[Ergodic capacity]{\label{fig:3a}\includegraphics[scale=0.5]{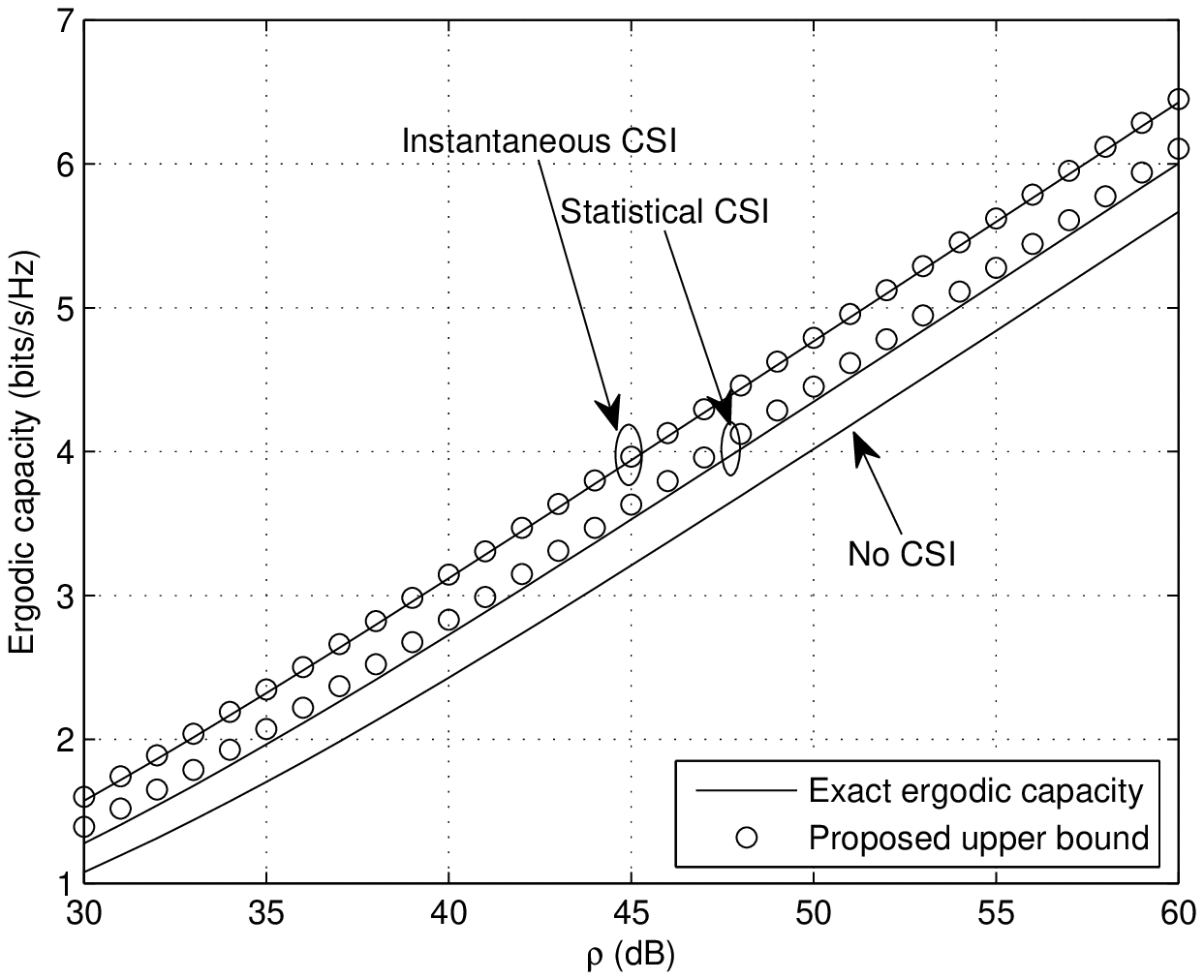}}
  \caption{Performance comparison of different cases when $N=3$, $\lambda=0.5$ and $\sigma=0.8$.}\label{fig:comparison of different CSI cases}
\end{figure}

Fig. \ref{fig:2a} and Fig. \ref{fig:3a} show achievable outage probability and ergodic capacity performances of three different cases, i.e.,instantaneous CSI, statistical CSI and no CSI, when $N=3$, $\lambda=0.5$ and $\sigma=0.8$. It can be observed from \ref{fig:2a} that the proposed lower bounds are very tight across the entire range of SNR, and the high SNR approximations are quite accurate, especially in the high SNR regime. In addition, we see that full diversity of $N$ is achieved with instantaneous CSI, yet the diversity order remains one with only statistical CSI or no CSI, which is consistent with our analytical results presented in Theorem \ref{theorem:2} and Theorem \ref{theorem:imCSI AF high SNR}. Similarly, Fig. \ref{fig:3a} indicates that the proposed analytical upper bounds in Theorem \ref{theorem:77} and Theorem \ref{theorem:imcsi ergodic} are sufficiently tight across the entire SNR region of interest. Finally, both figures demonstrate the intuitive results that the system performance improves with the available amount of CSI at the relay.

\begin{figure}[!ht]
  \centering
  \subfigure[Outage probability]{\label{fig:2a1}\includegraphics[scale=0.5]{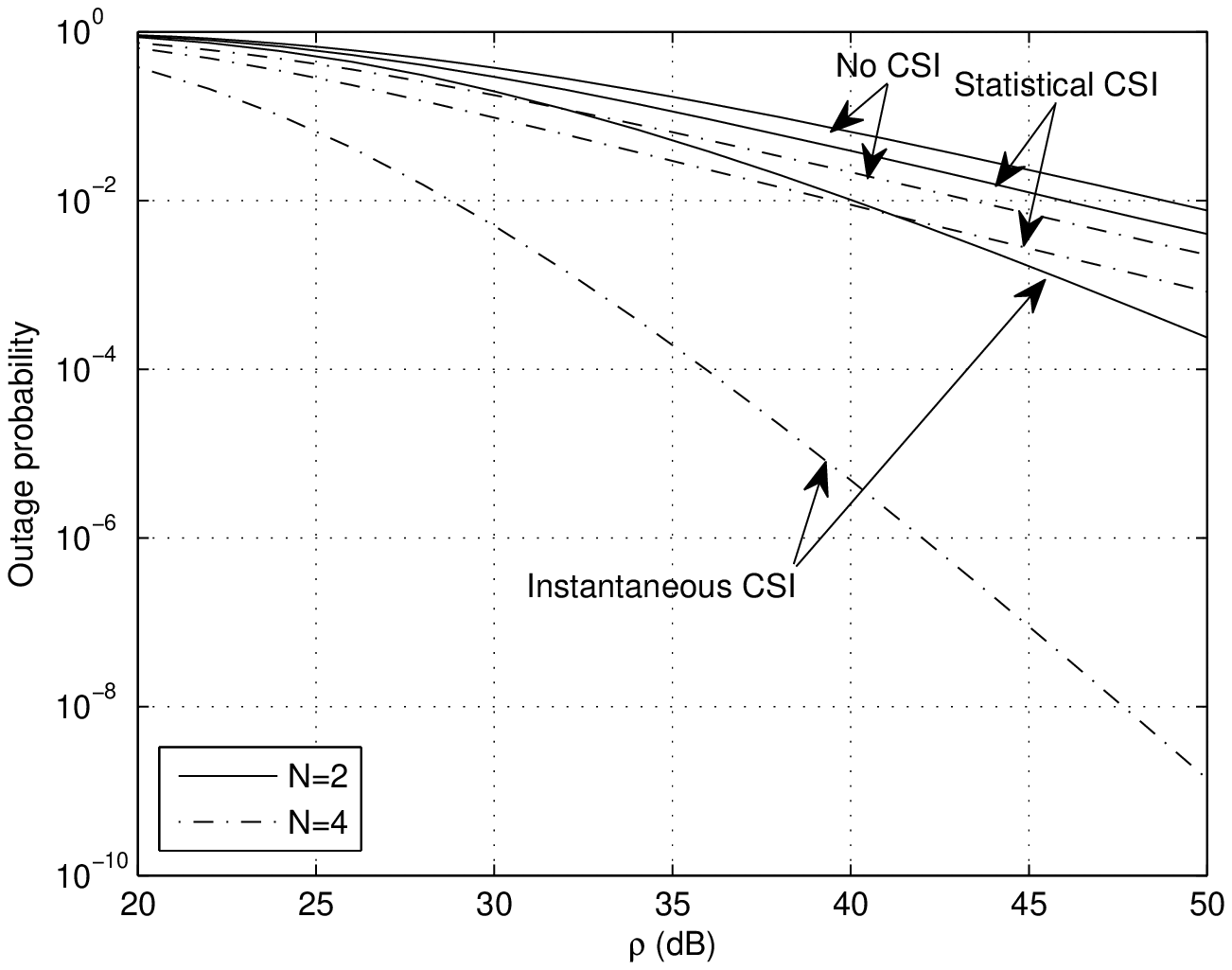}}
  \subfigure[Ergodic capacity]{\label{fig:3a1}\includegraphics[scale=0.5]{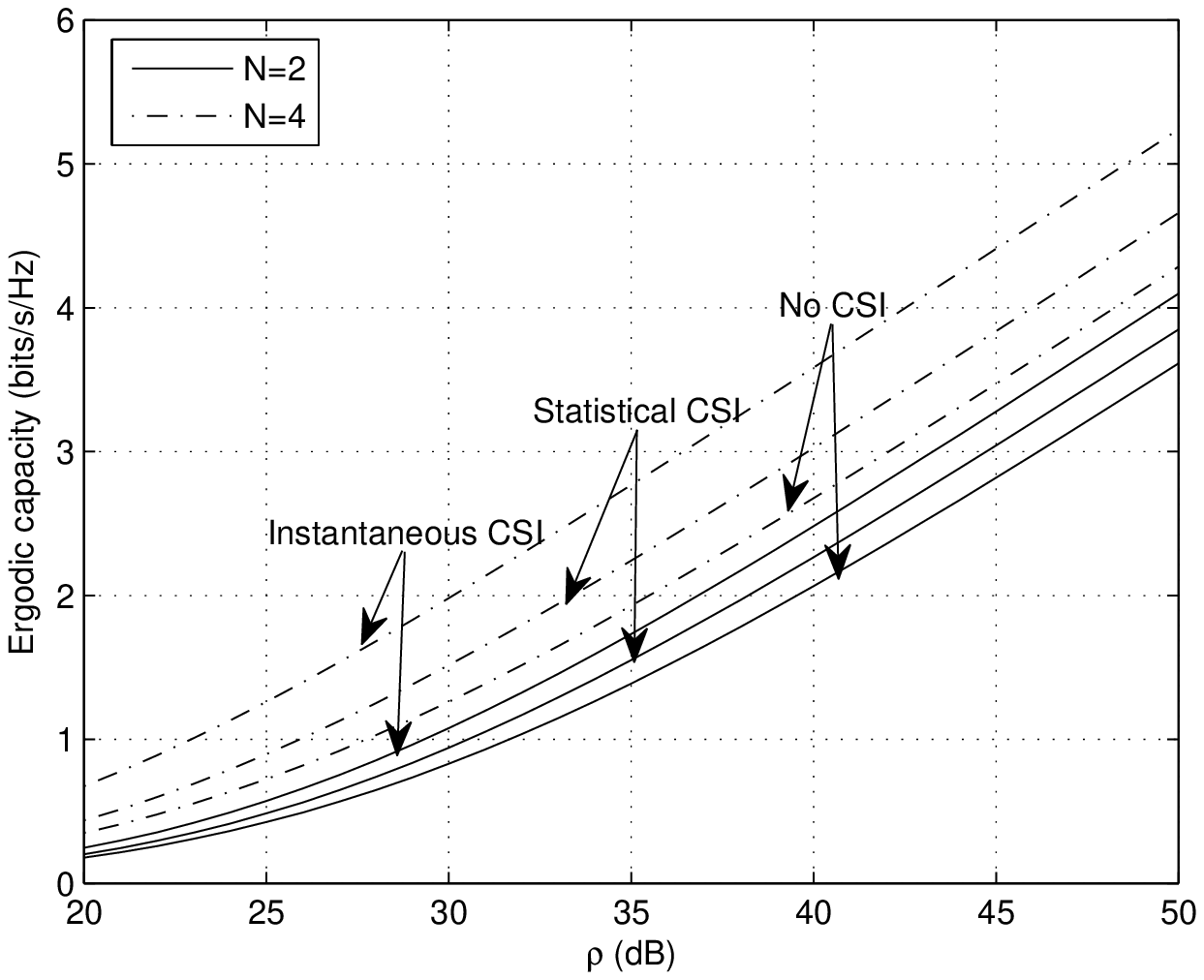}}
  \caption{Impact of $N$ on the system performance when $\lambda=0.5$ and $\sigma=0.8$.}\label{fig:impact of N on system performance}
\end{figure}

Fig. \ref{fig:impact of N on system performance} investigates the impact of $N$ on the system performance when $\lambda=0.5$ and $\sigma=0.8$. Intuitively, we observe that increasing the number of antennas significantly improves the outage probability and ergodic capacity performance. Moreover, the gain of increasing $N$ differs substantially for scenarios with different CSI assumptions. For instance, from Fig. \ref{fig:2a1}, the outage probability reduction is most pronounced for the instantaneous CSI scenario when $N$ increases from two to four. The reason is that, with instantaneous CSI, $N$ affects the diversity order, while for the other two scenarios, $N$ only provides some coding gain, hence the overall benefit is relatively small. Similar observations can be made from Fig. \ref{fig:3a1} in terms of ergodic capacity gain.

\begin{figure}[htbp]
    \centering
    \includegraphics[scale=0.55]{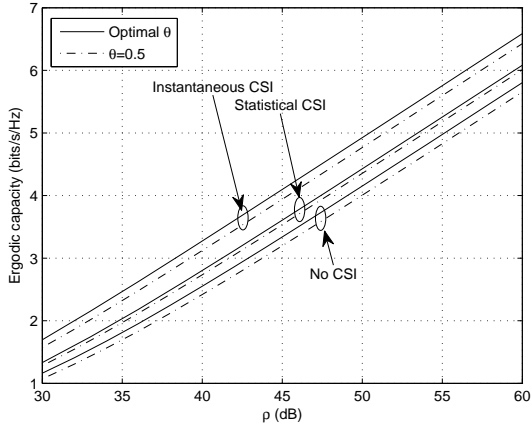}
    \caption{Impact of $\theta$ with $N=3$, $\rho=30$ dB, $\lambda=0.5$, $\sigma=0.8$, $d_1=d_2=3$ m.}\label{fig:fig41}
    \end{figure}

{In Fig. 2 and Fig. 3, a fixed power splitting ratio is used. Since the power splitting is a key design parameter, it is of interest to investigate its impact on the system performance. Fig. \ref{fig:fig41} illustrates the effect of optimal power splitting ratio on the system performance. The optimal power splitting ratio is obtained via numerical method based on the analytical expressions presented in Theorem 4 and 7. Interestingly, we observe that the performance gap between the ergodic capacity curves associated with the optimal $\theta$ and those with fixed $\theta=0.5$ is rather insignificant, especially for the case with statistical CSI.}

\begin{figure}[!ht]
  \centering
  \subfigure[$\rho=10$ dB]{\label{fig:Impact of correlation ins,cap,r=10}\includegraphics[scale=0.55]{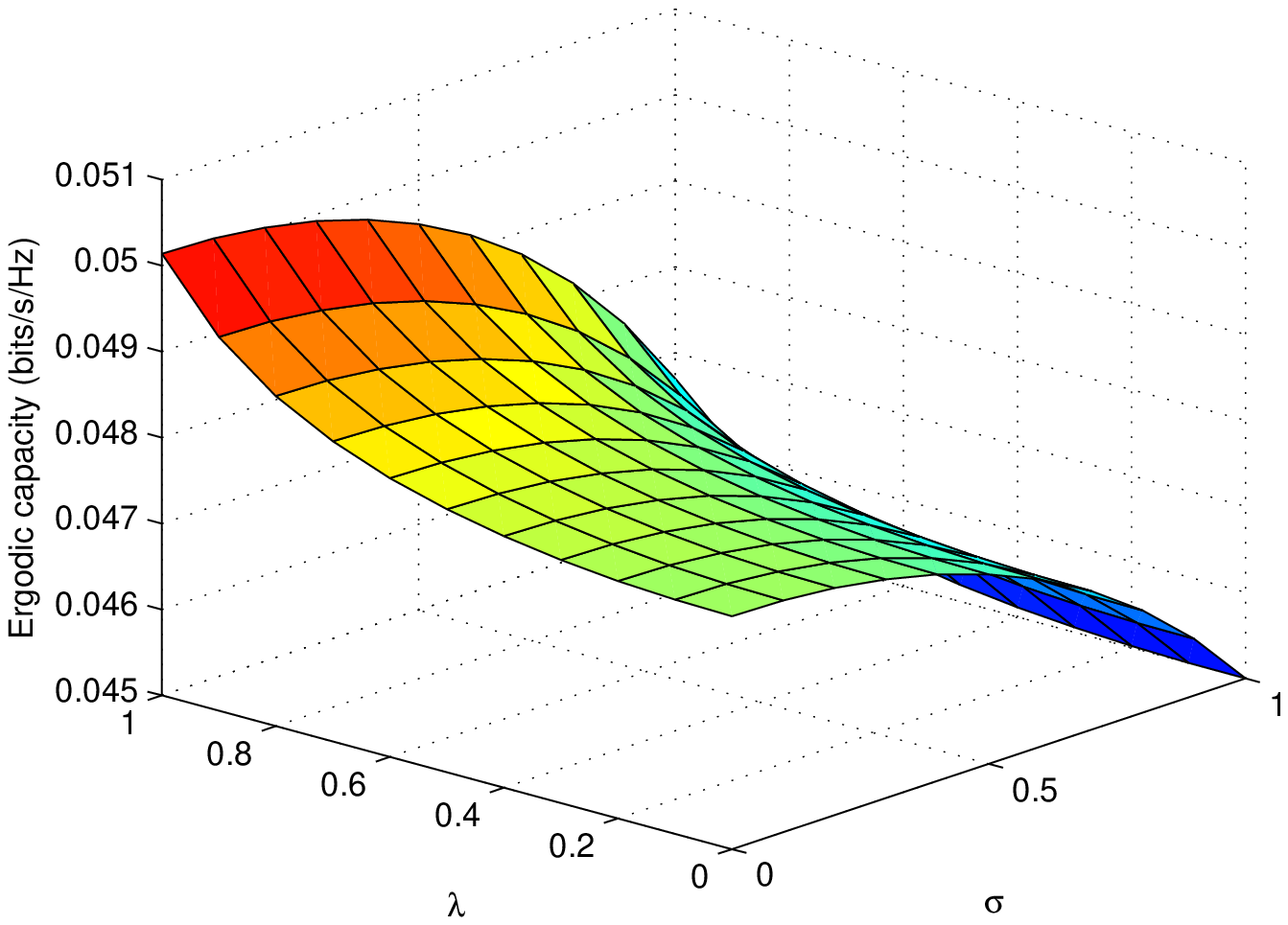}}
  \subfigure[$\rho=30$ dB]{\label{fig:Impact of correlation ins,cap,r=30}\includegraphics[scale=0.55]{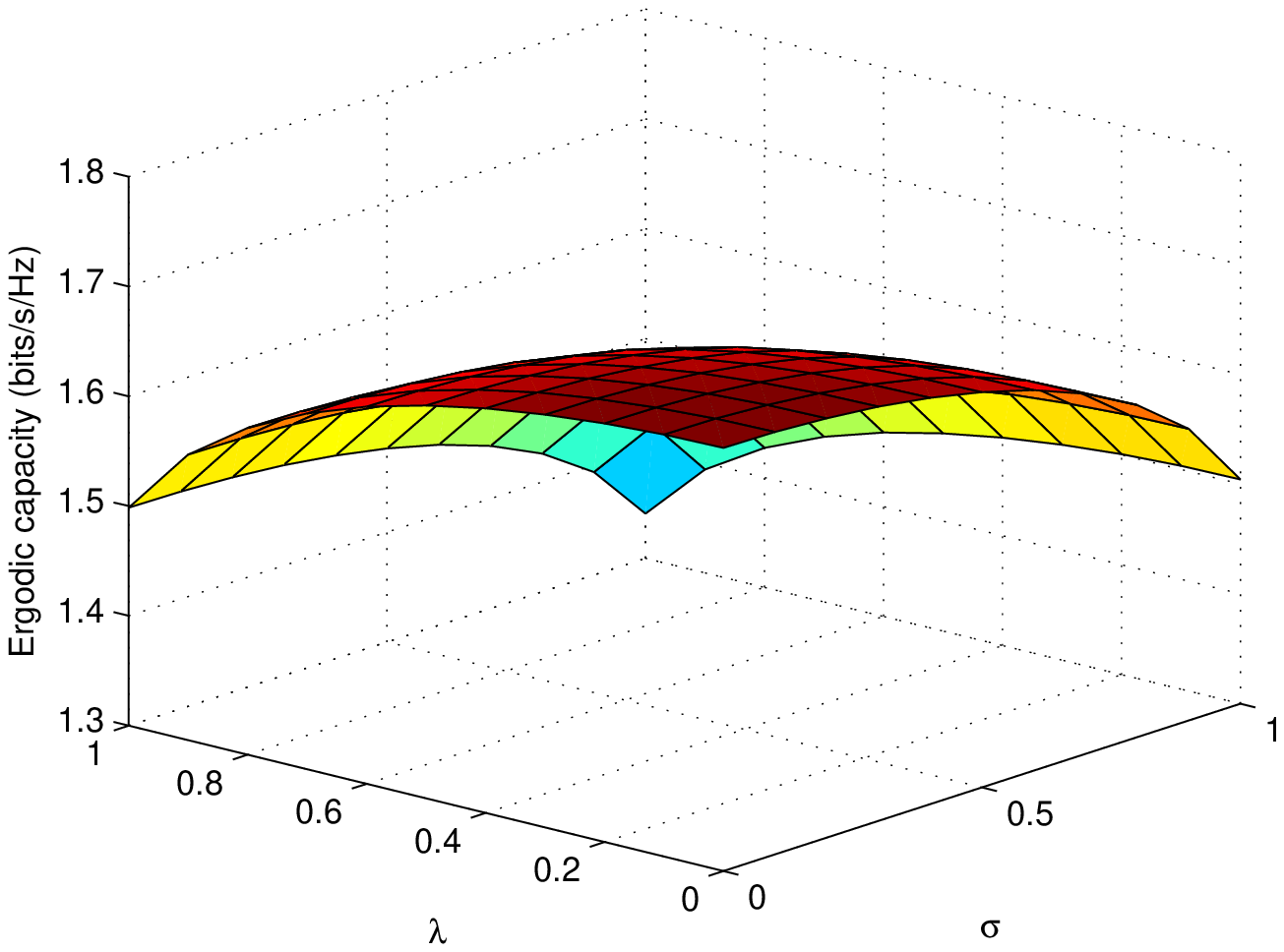}}
  \caption{Impact of antenna correlation on the ergodic capacity with instantaneous CSI.}\label{fig:Impact of correlation on capacity with instantaneous CSI}
\end{figure}

Fig. \ref{fig:Impact of correlation on capacity with instantaneous CSI} illustrates the impact of antenna correlation on the ergodic capacity with instantaneous CSI at different operating SNRs. For a fixed $\lambda$, the ergodic capacity is decreasing function with respect to $\sigma$ as shown in both figures, which indicates the transmit antenna correlation is always detrimental to the ergodic capacity, as analytically predicted in Proposition \ref{proposition:impact of correlation on ergodic capacity}. In contrast, for a fixed $\sigma$, we observe that the impact of receive antenna correlation on the ergodic capacity behaves quite different, for instance, when $\rho =10$ dB, stronger correlation increases the ergodic capacity, while when $\rho=30$ dB, stronger correlation decreases the ergodic capacity.

\begin{figure}[!ht]
  \subfigure[$\rho=10$ dB]{\label{fig:Impact of correlation st,cap,r=10}\includegraphics[scale=0.55]{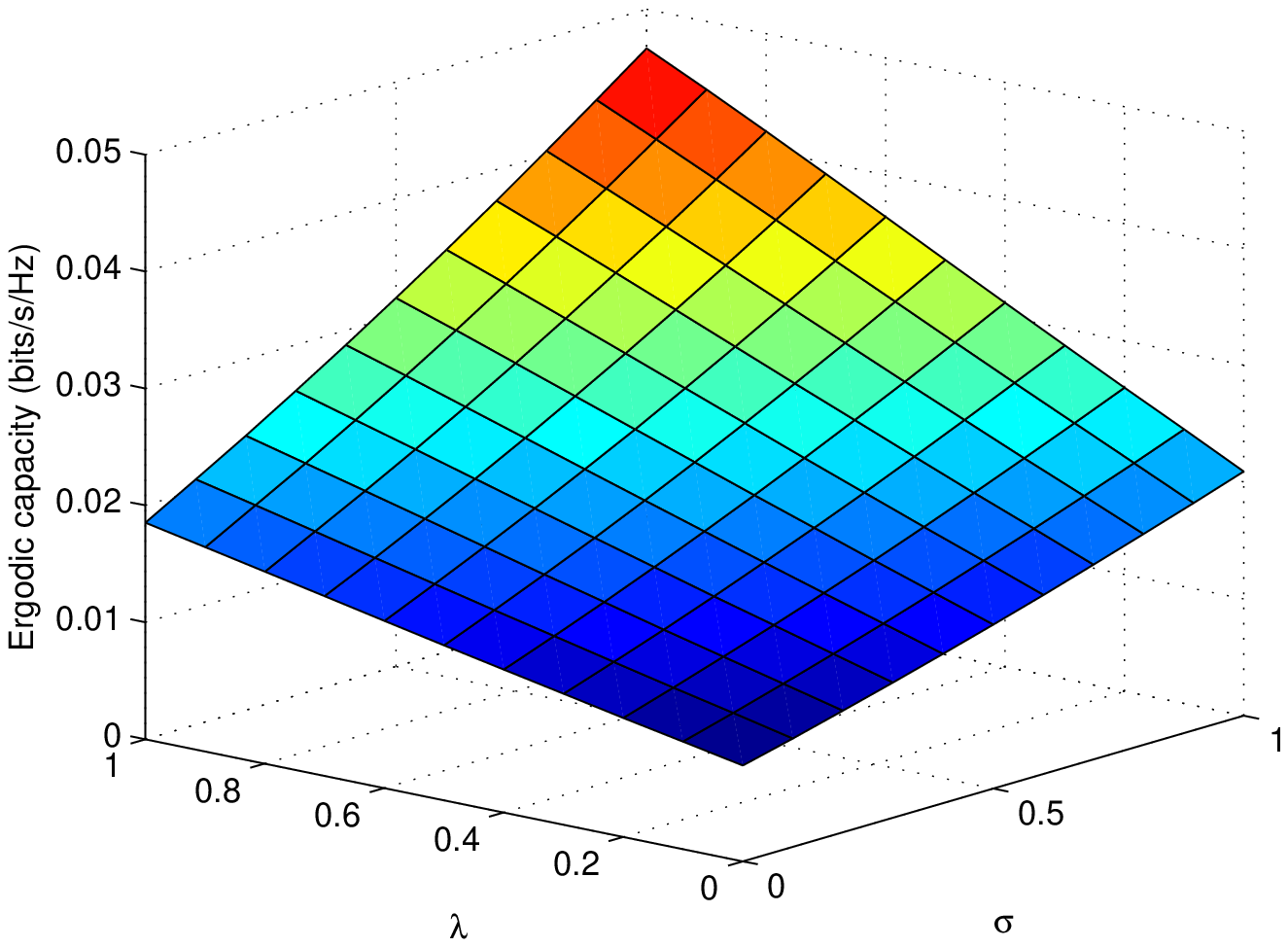}}
  \subfigure[$\rho=30$ dB]{\label{fig:Impact of correlation st,cap,r=30}\includegraphics[scale=0.55]{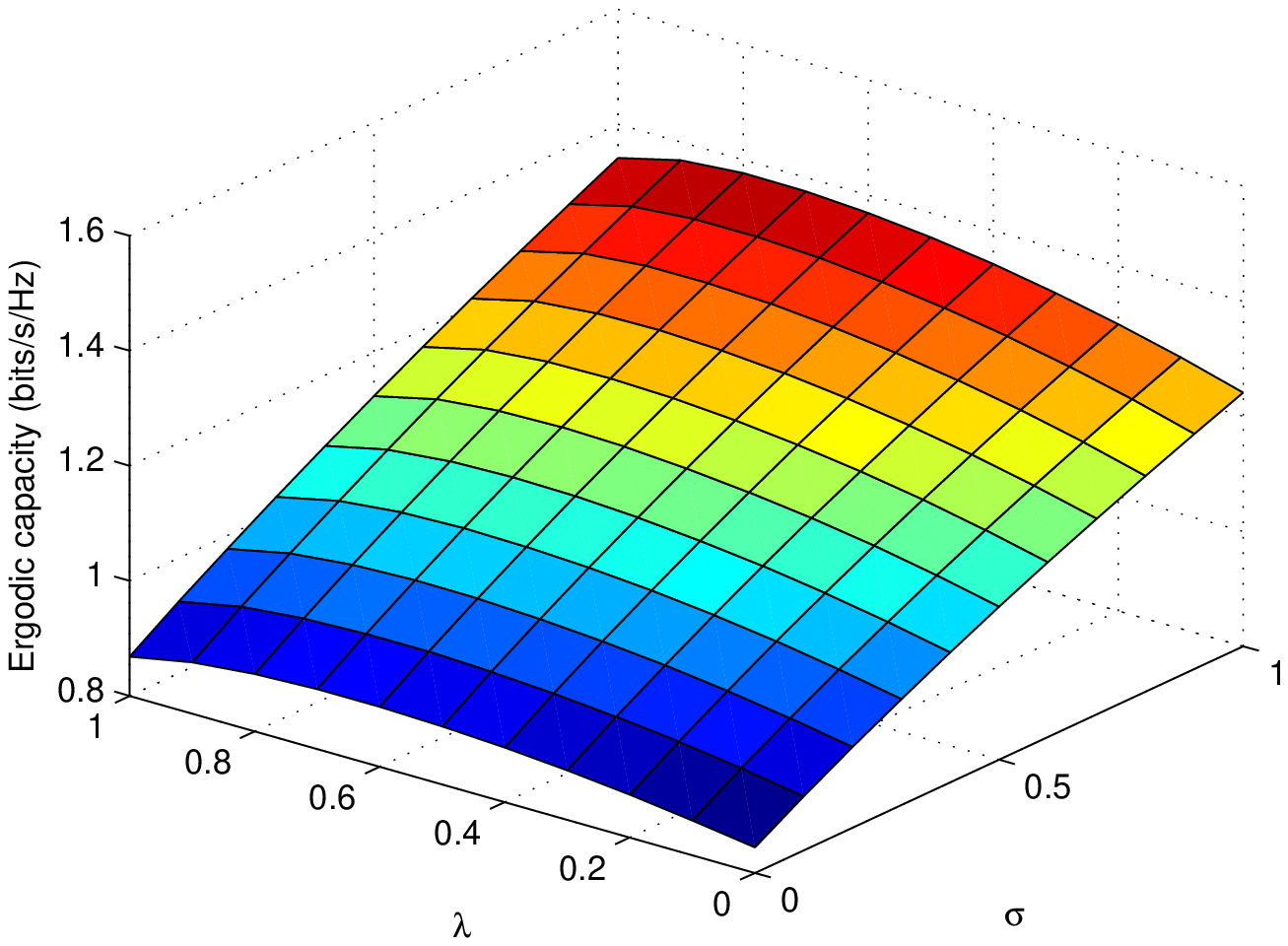}}
  \caption{Impact of antenna correlation on the ergodic capacity with statistical CSI.}\label{fig:Impact of correlation on capacity with statistical CSI}
\end{figure}

Fig. \ref{fig:Impact of correlation on capacity with statistical CSI} investigates the impact of antenna correlation on the ergodic capacity with statistical CSI. We observe that the transmit antenna correlation is always beneficial to the ergodic capacity, which is in consistent with the conclusion drawn in Proposition \ref{proposition impact of correlaion in statistical CSI}. As for the impact of receive antenna correlation, we observe it boosts the ergodic capacity at low SNRs which is similar to the scenario with instantaneous CSI. However, at high SNRs, we see a rather surprising behavior that the ergodic capacity first improves and then degrades when the correlation increases, indicating that there exists an optimal correlation point where the ergodic capacity is maximized.

\section{Conclusion}\label{section conclusion}
We have investigated the impact of antenna correlation and relay CSI on the performance of dual-hop wireless powered multi-antenna relay systems. Specifically, for both the instantaneous CSI and statistical CSI scenarios, we derived closed-form analytical expressions for the outage probability as well as simple high SNR approximations. In addition, we presented closed-form upper bound for the ergodic capacity, and characterized the impact of antenna correlation on the ergodic capacity. Our findings suggest that, the availability of CSI at the relay has a critical impact on the system performance. With instantaneous CSI, full diversity order can be achieved, while unit diversity order can be achieved with only statistical CSI. In addition, it is shown that the antenna correlation does not affect the achievable diversity order. Moreover, the impact of antenna correlation on the ergodic capacity depends on the available CSI and operating SNR, for instance, transmit antenna correlation is detrimental with instantaneous CSI while it is beneficial with only statistical CSI; receive antenna correlation improves ergodic capacity at lower SNRs, while it may degrade the ergodic capacity at moderate or high SNRs.

\appendices
\section{Proof of Theorem \ref{optimal scheme}}\label{appendix:optimal scheme}
To maximize the ergodic capacity, the optimization problem can be formulated as
\begin{align}
&\underset{{\bf{w}}_r, {{\bf{w}}_t}}\max~~{{\frac{1}{2}E}}\left\{\log_2\left(1+\frac{w^2\frac{(1-\theta)P_s}{d_1^\tau d_2^\tau}\left|{\bf{h}}_2{\bf{w}}_t{\bf{w}}_r{\bf{h}}_1\right|^2}{w^2\frac{\left|{\bf{h}}_2{\bf{w}}_t\right|^2N_0}{d_2^\tau}+N_0}\right)\right\}.\nonumber\\
&s.t. ~~{w}^2\frac{(1-\theta)P_s}{d_1^\tau}\left|{\bf{w}}_r{\bf{h}}_1\right|^2+{w}^2N_0={P_r}.
\end{align}
Since ${\bf{h}}_2{\bf{w}}_t$ and ${\bf{w}}_r{\bf{h}}_1$ are decoupled, the optimal ${\bf{w}}_r$ and ${{\bf{w}}_t}$ can be separately handled. As such, we first fix ${{\bf{w}}_r}$, hence, the original problem becomes
\begin{align}\label{P1}
\underset{{\bf{w}}_t}\max~~{{E}}\left\{\log_2\left(1+\frac{1}{A+B\frac{1}{\left|{\bf{h}}_2{\bf{w}}_t\right|^2}}\right)\right\},
\end{align}
where $A=\frac{d_1^\tau}{(1-\theta)\rho\left|{\bf{w}}_r{\bf{h}}_1\right|^2}$ and $B=\frac{d_1^\tau d_2^\tau}{\eta\theta\rho\left\|{\bf{h}}_1\right\|^2}+\frac{d_1^{2\tau} d_2^\tau}{\eta\theta(1-\theta)\rho^2\left\|{\bf{h}}_1\right\|^2\left|{\bf{w}}_r{\bf{h}}_1\right|^2}$.

Since ${\bf{h}}_2={\bf{h}}_{w2}{\bf{R}}_t^\frac{1}{2}$ and ${\bf{R}}_t^{\frac{1}{2}}={\bf{U}}_t{\bf{\Sigma}}_t^{\frac{1}{2}}{\bf{U}}_t^H$, (\ref{P1}) can be rewritten as
\begin{align}
\underset{{\bf{w}}_t}\max~~{{E}}\left\{\log_2\left(1+\frac{1}{A+\frac{B}{{\bf{h}}_{w2}{\bf{U}}_t{\bf{\Sigma}}_t^{\frac{1}{2}}{\bf{U}}_t^H{\bf{W}}_t{\bf{U}}_t{\bf{\Sigma}}_t^{\frac{1}{2}}{\bf{U}}_t^H{\bf{h}}_{w2}^H}}\right)
\right\},
\end{align}
where ${\bf{W}}_t={\bf{w}}_t{{\bf{w}}_t}^H$.
Noticing that ${\bf{h}}_{w2}$ is unitary invariant, by taking ${\bf{U}}_t^H{\bf{W}}_t{\bf{U}}_t=\widetilde{{\bf{W}}}_t$, we reformulate the problem as
\begin{align}
&\underset{\widetilde{\bf{w}}_t}\max~~{{E}}\left\{\log_2\left(1+\frac{1}{A+B\frac{1}{{\bf{h}}_{w2}{\bf{\Sigma}}_t^{\frac{1}{2}}\widetilde{\bf{W}}_t{\bf{\Sigma}}_t^{\frac{1}{2}}{\bf{h}}_{w2}^H}}\right)\right\}\nonumber\\
&=\underset{\widetilde{\bf{w}}_t}\max~C\left(\widetilde{\bf{W}}_t\right).
\end{align}

Let $\overline{\bf{W}}_t=\frac{1}{2}\left(\widetilde{\bf{W}}_t+{\bf{\Pi}}_k^H\widetilde{\bf{W}}_t{\bf{\Pi}}_k\right)$, where ${\bf{\Pi}}_k$, $1 \leqslant k \leqslant N$ is the $N\times N$ diagonal matrix with the $k$th diagonal element being $-1$, and all other diagonal elements being $1$. Now, noticing that $g({\bf{X}})={\bf{h}}_{w2}{\bf{\Sigma}}_t^{\frac{1}{2}}{\bf{X}}{\bf{\Sigma}}_t^{\frac{1}{2}}{\bf{h}}_{w2}^H$ is a concave function with respect to ${\bf{X}}$ and $h(x)=E\left\{\log_2\left(1+\frac{1}{A+B\frac{1}{x}}\right)\right\}$ is a concave and monotonic increasing function with respect to $x$ for any $A>0, B>0$, then, according to the composition theorem \cite[Ch. 3]{S.Boyd}, we establish that $C({\bf{W}})$ is a concave function. Therefore, we have
\begin{align}\label{ineq}
C\left(\overline{\bf{W}}_t\right)\geqslant\frac{1}{2}C\left(\widetilde{\bf{W}}_t\right)+\frac{1}{2}C\left({\bf{\Pi}}_k^H\widetilde{\bf{W}}_t{\bf{\Pi}}_k\right).
\end{align}
Now, since ${{\bf{\Pi}}_k}{\bf{\Sigma}}_t^{\frac{1}{2}}{\bf{\Pi}}_k^H={\bf{\Sigma}}_t^{\frac{1}{2}}$ and ${\bf{h}}_{w2}$ is unitary invariant, we have
\begin{align}\label{piwpi}
\hspace{-0.2cm}&C\left({\bf{\Pi}}_k^H\widetilde{\bf{W}}_t{\bf{\Pi}}_k\right)\notag\\
&={{E}}\left\{\log_2\left(1+\frac{1}{A+\frac{B}{{\bf{h}}_{w2}{\bf{\Sigma}}_r^{\frac{1}{2}}{\bf{\Pi}}_k^H\widetilde{\bf{W}}_t{\bf{\Pi}}_k{\bf{\Sigma}}_r^{\frac{1}{2}}{{\bf{h}}_{w2}^{H}}}}\right)\right\}\nonumber\\
\hspace{-0.2cm}&={E}\left\{\log_2\left(1+\frac{1}{A+\frac{B}{{{\bf{h}}_{w2}}{{\bf{\Pi}}_k}{\bf{\Sigma}}_r^{\frac{1}{2}}{\bf{\Pi}}_k^H\widetilde{\bf{W}}_t{\bf{\Pi}}_k{\bf{\Sigma}}_r^{\frac{1}{2}}{\bf{\Pi}}_k^H
{{\bf{h}}_{w2}^{H}}}}\right)\right\}\nonumber\\
\hspace{-0.2cm}&={{E}}\left\{\log_2\left(1+\frac{1}{A+\frac{B}{{{\bf{h}}_{w2}}{\bf{\Sigma}}_r^{\frac{1}{2}}\widetilde{\bf{W}}_t{\bf{\Sigma}}_r^{\frac{1}{2}}{{\bf{h}}_{w2}^{H}}}}\right)\right\}\notag\\
&=C\left(\widetilde{\bf{W}}_t\right).
\end{align}
Hence, we have
\begin{align}
C\left(\overline{\bf{W}}_t\right)\geqslant C\left(\widetilde{\bf{W}}_t\right),
\end{align}
where the equality holds if and only if $\overline{\bf{W}}_t=\widetilde{\bf{W}}_t$, indicating that the optimal $\widetilde{\bf{W}}_t$ is a diagonal matrix. Therefore, the optimal ${\bf{W}}_t$ can be expressed as ${\bf{W}}_t^{opt}={\bf{U}}_t{\bf{\Sigma}}_w{\bf{U}}_t^H$, where ${\bf{\Sigma}}_w$ is a diagonal matrix. Since ${\bf{W}}_t$ is a Rank-1 matrix, there is only one nonzero diagonal element in ${\bf{\Sigma}}_w$ which equals to one. To this end, it is easy to show that the 
%
optimal ${\bf{w}}_t$ is the eigenvector associated with the biggest eigenvalue of ${\bf{R}}_t$. Using the same method, the optimal ${\bf{w}}_r$ can be obtained.

\section{Proof of Theorem \ref{theorem:1}}\label{appendix:theorem:1}
Noticing that the end-to-end SNR (\ref{SNR 1}) can be upper bounded as
\begin{align}
\gamma_{\sf I} = \frac{\frac{\eta\theta(1-\theta)\rho^2}{d_1^{2\tau}}\left\|{\bf{h}}_2\right\|^2\left\|{\bf{h}}_1\right\|^4}{\frac{\eta\theta\rho}{d_1^\tau d_2^\tau}\left\|{\bf{h}}_2\right\|^2\left\|{\bf{h}}_1\right\|^2+\frac{(1-\theta)\rho}{d_1^\tau}\left\|{\bf{h}}_1\right\|^2}.
\end{align}
Then, (\ref{outage}) can be rewritten as
\begin{align}\label{outage:lower}
P_{I}^{low}=1-\int_{d/c}^{\infty}f_{\left\|{\bf{h}}_1\right\|^2}(x)\bar{F}_{\left\|{\bf{h}}_2\right\|^2}\left(\frac{a}{cx-d}\right)dx.
\end{align}
To solve the integral, we look into the PDFs of $\left\|{\bf{h}}_1\right\|^2$ and $\left\|{\bf{h}}_2\right\|^2$ and we have

\begin{align}
\left\|{\bf{h}}_1\right\|^2&=\left| {{\bf{h}}_{w1}}^H {{\bf{R}}_r} {{\bf{h}}_{w1}}\right|\nonumber\\
&=\left| {{\bf{h}}_{w1}}^H {\bf{U}}_r {\bf{\Sigma}}_r {\bf{U}}_r^H {{\bf{h}}_{w1}}\right|\nonumber\\
&=\sum_{i=1}^{N}\lambda_i \left|{\tilde{h}_{w1i}}\right|^2.
\end{align}
Since $\left|{\tilde{h}_{w1i}}\right|^2$ follows the exponential distribution, the PDF of $\left\|{\bf{h}}_1\right\|^2$ is the sum of $N$ weighted exponential random variables. With the help of\cite{Y.D.Yao}, the PDF of $\left\|{\bf{h}}_1\right\|^2$ can be written as
\begin{align}\label{pdf:coh1}
{f_{\left\|{\bf{h}}_1\right\|^2}(x)={\sum_{i=1}^N\prod_{\begin{subarray}{l}j=1\\j\ne i\end{subarray}}^N\frac{\lambda_i^{N-2}}{\lambda_i-\lambda_j}e^{-\frac{x}{\lambda_i}}}}.
\end{align}

Similarly, the PDF of $\left\|{\bf{h}}_2\right\|^2$ is given by
\begin{align}\label{pdf:coh2}
{f_{\left\|{\bf{h}}_2\right\|^2}(x)={\sum_{m=1}^N\prod_{\begin{subarray}{l}n=1\\n\ne m\end{subarray}}^N\frac{\sigma_m^{N-2}}{\sigma_m-\sigma_n}e^{-\frac{x}{\sigma_m}}}}.
\end{align}

Then, the right tail probability of $\left\|{\bf{h}}_2\right\|^2$ can be obtained as
\begin{align}\label{rtail:h2}
\bar{F}_{\left\|{\bf{h}}_2\right\|^2}\left(x\right) &= \int_{x}^\infty{\sum_{m=1}^N\prod_{\begin{subarray}{l}n=1\\n\ne m\end{subarray}}^N\frac{\sigma_m^{N-2}}{\sigma_m-\sigma_n}e^{-\frac{z}{\sigma_m}}}dz\\
&=\sum_{m=1}^N\prod_{\begin{subarray}{l}n=1\\n\ne m\end{subarray}}^N\frac{\sigma_m^{N-1}}{\sigma_m-\sigma_n}e^{-\frac{x}{\sigma_m}}.\label{fh2}
\end{align}

Substituting (\ref{rtail:h2}) and (\ref{pdf:coh1}) into (\ref{outage:lower}), the lower bound of the outage probability can be alternatively expressed as
\begin{align}
P_{I}^{low} = 1-\sum_{i=1}^N\prod_{\begin{subarray}{l}j=1\\j\ne i\end{subarray}}^N\frac{\lambda_i^{N-2}}{\lambda_i-\lambda_j}\sum_{m=1}^N\prod_{\begin{subarray}{l}n=1\\n\ne m\end{subarray}}^N\frac{\sigma_m^{N-1}}{\sigma_m-\sigma_n}\int_{d/c}^\infty e^{\frac{x}{\lambda_i}}e^{\frac{-a}{(cx-d)\sigma_m}}dx.
\end{align}
Making a change of variable $t=cx-d$ and with the help of \cite[Eq. (8.432.7)]{Tables}, the desired result can be obtained.

\section{Proof of Theorem \ref{theorem:2}}\label{appendix:theorem:2}
To prove Theorem \ref{theorem:2}, we need the following lemma:
\begin{lemma}\label{le1}
For an arbitrary $n$-dimensional square matrix $\bf{R}$ with eigenvalues $r_j$, $j\in \left[1, n\right]$ with decreasing order, i.e $r_{j_1} \geqslant r_{j_2}$ if $j_1<j_2$. We denote ${\bf{A}}({\bf{R}})_{i,j}=r_j^{i-1}$, $\det^{n,m}({\bf{A}}({\bf{R}}))$ represents the determinant of ${\bf{A}}({\bf{R}})$ with the $n$th row and $m$th column removed, then
\begin{equation}\label{lemma1}
\sum_{m=1}^n(-1)^{m+n}\sigma_m^{n-l-1}\det {}^{n,m}\left({\bf{A}}({\bf{R}})\right)=0
\end{equation}
for any real number $l \in [1,n]$.

\proof
We observe that (\ref{lemma1}) can be written as a determinant. And if $0< l < N$, the determinant becomes zero since the determinant contains two identical rows.
\endproof

\end{lemma}

To proceed with the proof of Theorem \ref{theorem:2}, as $\rho \rightarrow \infty$, the end-to-end SNR can be tightly upper bounded as
\begin{align}\label{upper bound}
\gamma_{\sf I} < \min\left\{\frac{(1-\theta)\rho\left\|{\bf{h}}_1\right\|^2}{d_1^\tau} , \frac{\eta\theta \rho}{d_1^\tau d_2^\tau}\left\|{\bf{h}}_1\right\|^2\left\|{\bf{h}}_2\right\|^2\right\}.
\end{align}

After some tedious manipulations, the outage probability can be rewritten as
\begin{multline}
P_{I}^{\infty} = P\left\{\left(\left\|{\bf{h}}_1\right\|^2 < \frac{\gamma_{\sf th}d_1^\tau}{(1-\theta)\rho}\right)\right\}+\\
P\left\{\left(\left\|{\bf{h}}_2\right\|^2 < \frac{d_1^\tau d_2^\tau\gamma_{\sf th}}{\eta\theta\rho\left\|{\bf{h}}_1\right\|^2}\right)\bigcup \left(\left\|{\bf{h}}_1\right\|^2 > \frac{\gamma_{\sf th}d_1^\tau}{(1-\theta)\rho}\right)\right\},
\end{multline}
%

Note that $\left\|{\bf{h}}_1\right\|^2$ and $\left\|{\bf{h}}_2\right\|^2$ are hypoexponential random variables, using the similar method as in \cite{R.Louie}, the outage probability can be written in the integral form as
\begin{multline}\label{pout}
P_{I}^{\infty}=1-\int_{\frac{\gamma_{\sf th} d_1^\tau}{(1-\theta)\rho}}^\infty \frac{-\det^{N,i}\left({\bf{A}}({\bf{R}}_r)\right)}{\Delta({\bf{R}}_t)\Delta({\bf{R}}_r)}\sum_{i=1}^N(-1)^{i+N}\lambda_i^{N-2}\times\\
{\sum_{m=1}^N(-1)^{m+N}\sigma_m^{N-l-1}\det {}^{N,m}\left({\bf{A}}({\bf{R}}_t)\right)}e^{-\frac{x}{\lambda_i}}e^{-\frac{d_1^\tau d_2^\tau \gamma_{\sf th}}{\eta\theta\rho\sigma_m x}}dx,
\end{multline}
where $\Delta({\bf{R}})$ represents the Vandermonde determinant of the eigenvalues of ${\bf{R}}$.

In the high SNR regime, i.e., $\frac{1}{\rho}\rightarrow0$, the most significant term is the one having the minimum order of $\frac{1}{\rho}$. Expanding $e^{-\frac{d_1^\tau d_2^\tau \gamma_{\sf th}}{\eta\theta\rho\sigma_m x}}$ in (\ref{pout}) using Taylor expansion, i.e., $\sum_{l=0}^\infty\left(\frac{-d_1^\tau d_2^\tau \gamma_{\sf th}}{\eta\theta\rho\sigma_m x}\right)^l\frac{1}{l!}$, according to Lemma \ref{le1}, the integrand in (\ref{pout}) equals to zero when $0<l<N$. Also, we notice that (\ref{pout}) contains higher order of $\frac{1}{\rho}$ when $l > N$. Hence, it is sufficient to look at the cases with $l=0$ or $l=N$ and (\ref{pout}) can be approximated as

\begin{multline}
P_{I}^{\infty}\approx 1-I_0+\frac{(-1)^{N+1}}{N!}\sum_{i=1}^{N}\sum_{m=1}^N\prod_{\begin{subarray}{l}j=1\\j\ne i\end{subarray}}^N\frac{\lambda_i^{N-2}}{\lambda_i-\lambda_j}\times\\
\prod_{\begin{subarray}{l}n=1\\n\ne m\end{subarray}}^N{\frac{\sigma_m^{N-1}}{\sigma_m-\sigma_n}}\left(\frac{d_1^\tau d_2^\tau \gamma_{\sf th}}{\eta\theta\rho}\right)^N\int_{\frac{\gamma_{\sf th} d_1^\tau}{(1-\theta)\rho}}^\infty e^{-\frac{x}{\lambda_i}}x^{-N}dx,
\end{multline}
where

\begin{align}\label{I0}
I_0 = \sum_{i=1}^{N}\sum_{m=1}^N\prod_{\begin{subarray}{l}j=1\\j\ne i\end{subarray}}^N\frac{\lambda_i^{N-1}}{\lambda_i-\lambda_j}\prod_{\begin{subarray}{l}n=1\\n\ne m\end{subarray}}^N{\frac{\sigma_m^{N-1}}{\sigma_m-\sigma_n}}e^{-\frac{\gamma_{th}d_1^\tau}{(1-\theta)\rho\lambda_i}}.
\end{align}

Expanding $e^{-\frac{\gamma_{th}d_1^\tau}{(1-\theta)\rho\lambda_i}}$ in (\ref{I0}) using Taylor expansion and omitting the higher order items of $\frac{1}{\rho}$, (\ref{I0}) can be approximated as
\begin{align}
I_0 \approx 1+\sum_{i=1}^{N}\sum_{m=1}^N\prod_{\begin{subarray}{l}j=1\\j\ne i\end{subarray}}^N\frac{\lambda_i^{-1}}{\lambda_i-\lambda_j}\prod_{\begin{subarray}{l}n=1\\n\ne m\end{subarray}}^N{\frac{\sigma_m^{N-1}}{\sigma_m-\sigma_n}}\frac{\left(\frac{-\gamma_{\sf th}d_1^\tau}{(1-\theta)\rho}\right)^N}{N!}.\notag
\end{align}

Making a change of variable $t=\frac{x}{\lambda_i}$, and with the help of\cite[Eq. (8.350.2)]{Tables}, \cite[Eq. (8.352.8)]{Tables} and \cite[Eq. (8.214.1)]{Tables}, we have
\begin{multline}\label{PI_final}
P_{I}^{\infty}\approx 1-\frac{\left(\frac{d_1^\tau d_2^\tau \gamma_{\sf th}}{\eta\theta\rho}\right)^N}{N!(N-1)!}\sum_{i=1}^{N}\sum_{m=1}^N\prod_{\begin{subarray}{l}j=1\\j\ne i\  \end{subarray}}^N\frac{\lambda_i^{-1}}{\lambda_i-\lambda_j}
\prod_{\begin{subarray}{l}n=1\\n\ne m\end{subarray}}^N{\frac{\sigma_m^{-1}}{\sigma_m-\sigma_n}}\\
\times\left[{\sf Ei}\left(\frac{-\gamma_{\sf th}d_1^\tau}{(1-\theta)\rho\lambda_i}\right)+\sum_{m=0}^{N-2}\frac{(-1)^me^{\frac{-\gamma_{\sf th}d_1^\tau}
{(1-\theta)\rho\lambda_i}}m!}{\left(\frac{\gamma_{\sf th}d_1^\tau}{(1-\theta)\rho\lambda_i}\right)^{m+1}}\right]-I_0.
\end{multline}

Finally, expanding $e^{\frac{-\gamma_{\sf th}d_1^\tau}
{(1-\theta)\rho\lambda_i}}$ in (\ref{PI_final}) using Taylor expansion, applying\cite[Eq. (8.214.1)]{Tables}, and omitting the terms containing higher order of $\frac{1}{\rho}$, the desired result can be obtained.

\section{Proof of Theorem \ref{theorem:77}}\label{appendix:theorem:77}
Note that (\ref{C}) can be rewritten as
\begin{align}\label{capacity}
C_{I} = C_{\gamma_1}+C_{\gamma_2}-C_{\gamma_{r}},
\end{align}
where $C_{\gamma_1}=\frac{1}{2}E\left[\log_2(1+\gamma_1)\right]$, $C_{\gamma_2}=\frac{1}{2}E\left[\log_2(1+\gamma_2)\right]$, and $C_{\gamma r}=\frac{1}{2}E\left[\log_2(1+\gamma_1+\gamma_2)\right]$.
Exploiting the fact that $f(x,y) = \log_2{(1+e^x+e^y)}$ is a convex function with respect to x and y, $C_{\gamma_{r}}$ can be bounded as
\begin{align}
{C}_{\gamma r}\geqslant \hat{C}_{\gamma r} = \frac{1}{2}\log_2{\left(1+e^{E(\ln{\gamma_1})}+e^{E(\ln{\gamma_2})}\right)}.
\end{align}

Hence, the ergodic capacity in (\ref{capacity}) can be upper bounded as
\begin{align}\label{Ebound:1}
C_{I}^{up}=C_{\gamma_1}+C_{\gamma_2}-\frac{1}{2}\log_2\left(1+e^{E(\ln{\gamma_1})}+e^{E(\ln{\gamma_2})}\right).
\end{align}

The remaining task is to compute $C_{\gamma_1}$, $C_{\gamma_2}$ and $\hat{C}_{\gamma r}$. Utilizing the same method as in \cite{G.Zhu.two}, $C_{\gamma_i}$, $\gamma_i \in \left\{\gamma_1,\gamma_2\right\}$, can be derived as
\begin{align}\label{bothcoc1}
C_{\gamma_i} = \frac{1}{2\ln{2}}\int_0^\infty\frac{1-F_{\gamma_i}(x)}{1+x}dx.
\end{align}

\subsection{Computation of $C_{\gamma_1}$}
Given the PDF of $\left\|{\bf{h}}_1\right\|^2$ in (\ref{pdf:coh1}), CDF of $\gamma_1$ can be expressed as
\begin{align}\label{Fh1co}
F_{\gamma_1}(x)= 1 - \sum_{i=1}^N\prod_{\begin{subarray}{l}j=1\\j \ne i\end{subarray}}^N\frac{\lambda_i^{N-1}}{\lambda_i-\lambda_j}e^{-\frac{d_1^\tau x}{(1-\theta)\rho\lambda_i}}.
\end{align}

Substituting (\ref{Fh1co}) into (\ref{bothcoc1}), with the help of \cite[Eq. (8.350.2)]{Tables}, $C_{\gamma_1}$ can be computed as
\begin{align}\label{Cgamma1:1}
C_{\gamma_1} = -\frac{1}{2\ln{2}}\sum_{i=1}^N\prod_{\begin{subarray}{l}j=1\\j \ne i\end{subarray}}^N\frac{\lambda_i^{N-1}}{\lambda_i-\lambda_j}e^{\frac{d_1^\tau}{(1-\theta)\rho\lambda_i}}{\sf Ei}\left(-\frac{d_1^\tau}{(1-\theta)\rho\lambda_i}\right),
\end{align}

\subsection{Computation of $C_{\gamma_2}$}
Similarly, the CDF of $\gamma_2$ can be expressed as
\begin{align}\label{h1cocgamma2}
F_{\gamma_{2}}(x) = 1 - \int_0^\infty\bar{F}_{\left\|{\bf{h}}_2\right\|^2}\left(\frac{d_1^\tau d_2^\tau x}{\eta\theta\rho t^2}\right)f_{\left\|{\bf{h}}_1\right\|^2}(t)dt.
\end{align}
Given the $\bar{F}_{\left\|{\bf{h}}_2\right\|^2}\left(x\right)$ in (\ref{fh2}) and the PDF of $\left\|{\bf{h}}_1\right\|^2$ in (\ref{pdf:coh1}), with the help of \cite[Eq. (8.432.7)]{Tables}, (\ref{h1cocgamma2}) can be obtained as
\begin{align}\label{h1coFcgamma2}
F_{\gamma_{2}}(x) = 1-2\sum_{i=1}^N\sum_{m=1}^N\prod_{\begin{subarray}{l}j=1\\j \ne i\end{subarray}}^N\frac{\lambda_i^{N-1}}{\lambda_i-\lambda_j}\prod_{\begin{subarray}{l}m=1\\m \ne n\end{subarray}}^N\frac{\sigma_m^{N-1}}{\sigma_m-\sigma_n}\sqrt{\frac{d_1^\tau d_2^\tau x}{\eta\theta\rho\lambda_i\sigma_m}}K_{1}\left(2\sqrt{\frac{d_1 ^\tau d_2^\tau x}{\eta\theta\rho\lambda_i\sigma_m}}\right).
\end{align}
Then, substituting (\ref{h1coFcgamma2}) into (\ref{bothcoc1}), $C_{\gamma_2}$ can be obtained as
\begin{align}\label{cgamma2:0}
C_{\gamma_2} = \frac{1}{2\ln{2}}\sum_{i=1}^N\sum_{m=1}^N\prod_{\begin{subarray}{l}j=1\\j \ne i\end{subarray}}^N\frac{\lambda_i^{N-1}}{\lambda_i-\lambda_j}\prod_{\begin{subarray}{l}m=1\\m \ne n\end{subarray}}^N\frac{\sigma_m^{N-1}}{\sigma_m-\sigma_n}\int_0^\infty\frac{2\sqrt{\frac{d_1 ^\tau d_2^\tau x}{\eta\theta\rho\lambda_i}}K_{1}\left(2\sqrt{\frac{d_1 ^\tau d_2^\tau x}{\eta\theta\rho\lambda_i\sigma_m}}\right)}{1+x}dx.
\end{align}
With the help of \cite[Eq. (9.343)]{Tables} and \cite[Eq. (7.811.5)]{Tables}, (\ref{cgamma2:0}) can be expressed as
\begin{align}\label{Cgamma2:1}
C_{\gamma_2} = \frac{1}{2\ln{2}}\sum_{i=1}^N\sum_{m=1}^N\prod_{\begin{subarray}{l}j=1\\j \ne i\end{subarray}}^N\frac{\lambda_i^{N-1}}{\lambda_i-\lambda_j}\prod_{\begin{subarray}{l}m=1\\m \ne n\end{subarray}}^N\frac{\sigma_m^{N-1}}{\sigma_m-\sigma_n}G_{1 3}^{3 1}\left(\frac{d_1^\tau d_2^\tau}{\eta\theta\rho\lambda_i\sigma_m}\middle |_{0,1,0}^0\right).
\end{align}

\subsection{Computation of $E\left(\ln{\gamma_1}\right)$}\label{appendix:theorem:7:c}

The expectation of $\ln{\gamma_i}$, $\gamma_i\in\left\{\gamma_1,\gamma_2\right\}$, can be derived using following approach:
\begin{align}\label{Eln1:1}
E\left(\ln{\gamma_i}\right) = \frac{dE\left(\gamma_i^n\right)}{dn}\bigg |_{n=0}.
\end{align}

Note that the $n$-th moment of $\gamma_i\in\left\{\gamma_1,\gamma_2\right\}$ can be computed as
\begin{align}\label{Egamma1:1}
E\left(\gamma_i^n\right) = \int_0^\infty x^n f_{\gamma_i}(x)dx = n\int_0^\infty x^{n-1}\left(1-F_{\gamma_i}(x)\right)dx.
\end{align}

Substituting (\ref{Fh1co}) into (\ref{Egamma1:1}) with the help of \cite[Eq. (8.310.1)]{Tables}, we can obtain
\begin{align}\label{Egamma1final:1}
E\left(\gamma_1^n\right) = \Gamma(n+1)\sum_{i=1}^N\prod_{\begin{subarray}{l}j=1\\j \ne i\end{subarray}}^N\frac{\lambda_i^{N-1}}{\lambda_i-\lambda_j}\left(\frac{(1-\theta)\rho\lambda_i}{d_1^\tau}\right)^n.
\end{align}

Finally, substituting (\ref{Egamma1final:1}) into (\ref{Eln1:1}), $E\left(\ln{\gamma_1}\right)$ can be expressed as
\begin{align}\label{elngamma1}
E\left(\ln{\gamma_1}\right) = \sum_{i=1}^N\prod_{\begin{subarray}{l}j=1\\j \ne i\end{subarray}}^N\frac{\lambda_i^{N-1}}{\lambda_i-\lambda_j}\left(\psi(1)+\ln{\left(\frac{(1-\theta)\rho\lambda_i}{d_1^\tau}\right)}\right).
\end{align}

\subsection{Computation of $E\left(\ln{\gamma_2}\right)$}
Utilizing the CDF of $\gamma_2$ in (\ref{h1coFcgamma2}), following similar methods as the computation of $E\left(\ln{\gamma_1}\right)$ with the help of \cite[Eq. (6.561.16)]{Tables}, $E\left(\ln{\gamma_2}\right)$ can be obtained as
\begin{align}\label{Eln2:1}
E\left(\ln{\gamma_2}\right) = \sum_{i=1}^N\sum_{m=1}^N\prod_{\begin{subarray}{l}j=1\\j \ne i\end{subarray}}^N\frac{\lambda_i^{N-1}}{\lambda_i-\lambda_j}\prod_{\begin{subarray}{l}m=1\\m \ne n\end{subarray}}^N\frac{\sigma_m^{N-1}}{\sigma_m-\sigma_n}\left(2\psi(1)-\ln{\frac{d_1^\tau d_2^\tau}{\eta\theta\rho\lambda_i\sigma_m}}\right).
\end{align}

Finally, substituting (\ref{Cgamma1:1}), (\ref{Cgamma2:1}), (\ref{elngamma1}) and (\ref{Eln2:1}) into (\ref{Ebound:1}) yields the desired result.

\section{Proof of Proposition \ref{proposition:impact of correlation on ergodic capacity}}\label{appendix:impact of correlation on ergodic capacity}
We start by introducing the following essential definitions and lemma on majorization theory given in \cite[Ch. 3, Ch. 11]{A.W.Marshall}.
\begin{definition}\label{definition 2}
A real-valued function $\Psi({\bf{x}})$: $\mathbb{R}^n \rightarrow \mathbb{R}$ is said to be \emph{Schur-convex} if $\bf{x} \succ \bf{y} \Rightarrow \Psi(x) \geqslant \Psi(y)$ and is said to be \emph{Schur-concave} if $\bf{x} \succ \bf{y} \Rightarrow \Psi(x) \leqslant \Psi(y)$.
\end{definition}

\begin{definition}
Random variables $X_1, X_2, \cdots, X_n$ are exchangeable random variables if the joint distribution of $X_1, X_2, \cdots, X_n$ is invariant under permutations of its arguments.
\end{definition}

\begin{remark}\label{remark 1}
I.i.d random variables are exchangeable.
\end{remark}

\begin{lemma}\label{lemma 2}
If $X_1, X_2, \cdots, X_n$ are exchangeable random variables and $\varepsilon({\bf{x}})$ is continuous and convex, and
\begin{align}
\Psi\left(\mu_1,\mu_2,\cdots, \mu_n \right)=E\left\{\varepsilon\left(\sum_{i=1}^n\mu_iX_i\right)\right\}.
\end{align}
Then $\Psi$ is \emph{Schur-convex}. Similarly, if $\varepsilon({\bf{x}})$ is continuous and concave, $\Psi$ is \emph{Schur-concave}.
\end{lemma}

Define $x\triangleq\sum_{i=1}^N\lambda_i\left|{\tilde{h}_{w1i}}\right|^2$ and $y\triangleq\sum_{m=1}^N\sigma_m\left|{\tilde{h}_{w2m}}\right|^2$, then the instantaneous capacity can be expressed as
\begin{align}\label{varexy}
C(x,y)=\log_2\left(1+\frac{m_1n_1x^2y}{m_1x+n_1xy+1}\right).
\end{align}
where $m_1=\frac{(1-\theta)\rho}{d_1^\tau}$ and $n_1=\frac{\eta\theta\rho}{d_1^\tau d_2^\tau}$.

To investigate the impact of transmit antenna correlation, we first fix $x$, hence, it is easy to prove the second order derivative of $C(x,y)$ with respect to $y$ can be derived as
\begin{align}
\frac{d^2C(x,y)}{d y^2}=-\frac{m_1n_1^2x\left(2+m_1x+2n_1y\right)}{\left(1+n_1y\right)^2\left(1+m_1x+n_1y\right)^2\ln2}<0,
\end{align}
which indicates the $C(x,y)$ is a concave function with respect to $y$.

We now look into the impact of receive antenna correlation. In the high SNR regime, the instantaneous capacity can be accurately approximated by
\begin{align}\label{varexy1}
C(x,y)\approx C^h(x,y)=\log_2\left(\frac{m_1n_1x^2y}{m_1x+n_1xy+1}\right).
\end{align}
With fixed $y$, the second order derivative of $C^h(x,y)$ with respect to $x$ can be derived as
\begin{align}
\frac{d^2C^h(x,y)}{d x^2}=-\frac{2+4(n_1y+m_1)x+(m_1^2+n_1^2y^2+2m_1n_1y)x^2}{x^2\left(1+m_1x+n_1yx\right)^2\ln2}<0,
\end{align}
which indicates the $C^h(x,y)$ is a concave function with respect to $x$.

In the low SNR regime, the instantaneous capacity can be accurately approximated by
\begin{align}\label{varexy2}
C(x,y)\approx C^l(x,y)=\frac{m_1n_1x^2y}{m_1x+n_1xy+1}.
\end{align}
With fixed $y$, the second order derivative of $C^l(x,y)$ with respect to $x$ can be derived as
\begin{align}
\frac{d^2C^l(x,y)}{d x^2}=\frac{2m_1n_1y}{\left(1+m_1x+n_1yx\right)^3\ln2}>0,
\end{align}
which indicates the $C^l(x,y)$ is a convex function with respect to $x$.

To this end, invoking Lemma \ref{lemma 2} yields the desired result.

\section{Proof of Theorem \ref{theorem:imCSI AF outage}}\label{appendix:imCSI AF outage}
Based on the end-to-end SNR given in (\ref{inscsi snr}), the outage probability can be computed by
\begin{align}
P_{S}=P\left\{\frac{\frac{\eta\theta(1-\theta)\rho^2}{d_1^{2\tau}d_2^\tau}\left\|{\bf{h}}_1\right\|^2\lambda_1\left|\tilde{h}_{w11}\right|^2\sigma_1\left|\tilde{h}_{w21}\right|^2}{\frac{\eta\theta\rho\sigma_1}{d_1^\tau d_2^\tau}\left\|{\bf{h}}_1\right\|^2\left|\tilde{h}_{w21}\right|^2+\frac{(1-\theta)\rho}{d_1^\tau}\lambda_1\left|\tilde{h}_{w11}\right|^2+1}<\gamma_{\sf th}\right\}.
\end{align}
After some simple manipulation, the outage probability can be rewritten as
\begin{multline}
P_{S}=P\left\{\left\|{\bf{h}}_1\right\|^2\left|\tilde{h}_{w21}\right|^2<\frac{\frac{(1-\theta)\rho\lambda_1\gamma_{\sf th}}{d_1^\tau}\left|\tilde{h}_{w11}\right|^2+\gamma_{\sf th}}{\frac{\eta\theta(1-\theta)\rho^2\lambda_1\sigma_1}{d_1^{2\tau}d_2^\tau}
\left|\tilde{h}_{w11}\right|^2-\frac{\eta\theta\rho\gamma_{\sf th}\sigma_1}{d_1^\tau d_2^\tau}}
\cup \left|\tilde{h}_{w11}\right|^2>\frac{\gamma_{\sf th}d_1^\tau}{(1-\theta)\rho} \right\}\\+P\left\{\left|\tilde{h}_{w11}\right|^2<\frac{\gamma_{\sf th}d_1^\tau}{(1-\theta)\rho}\right\}.
\end{multline}
Noticing that
\begin{align}
\left\|{\bf{h}}_1\right\|^2=\sum_{i=2}^N\lambda_i\left|\tilde{h}_{w1i}\right|^2+\lambda_1\left|\tilde{h}_{w11}\right|^2,
\end{align}
the outage probability can be computed as
\begin{multline}\label{imCSI AF outage de}
P_{S}=P\left\{\left|\tilde{h}_{w11}\right|^2<\frac{\gamma_{\sf th}d_1^\tau}{(1-\theta)\rho}\right\}+\\
P\left\{\sum_{i=2}^N\lambda_i\left|\tilde{h}_{w1i}\right|^2<\frac{\frac{(1-\theta)\rho\lambda_1\gamma_{\sf th}}{d_1^\tau}\left|\tilde{h}_{w11}\right|^2+\gamma_{\sf th}}{\frac{\eta\theta(1-\theta)\rho^2\lambda_1\sigma_1}{d_1^{2\tau}d_2^\tau}
\left|\tilde{h}_{w11}\right|^2-\frac{\eta\theta\rho\gamma_{\sf th}\sigma_1}{d_1^\tau d_2^\tau}}\frac{1}{\left|\tilde{h}_{w21}\right|^2}-\lambda_1\left|\tilde{h}_{w11}\right|^2 \cup \left|\tilde{h}_{w11}\right|^2>\frac{\gamma_{\sf th}d_1^\tau}{(1-\theta)\rho}\right\}.
\end{multline}
Since $\sum_{i=2}^N\lambda_i\left|\tilde{h}_{w1i}\right|^2$ is a hyper-exponential random variable, and $\left|\tilde{h}_{w11}\right|^2$ is an exponential random variable, the desired result can be obtained after some algebraic manipulations.

\section{Proof of Theorem \ref{theorem:imCSI AF high SNR}}\label{appendix:imCSI AF high SNR}
In the high SNR regime, the outage probability can be tightly approximated as
\begin{align}
P_{S}\approx P\left\{\min\left\{\frac{\eta\theta\rho\sigma_1}{d_1^\tau d_2^\tau}\left\|{\bf{h}}_1\right\|^2\left|\tilde{h}_{w21}\right|^2,\frac{(1-\theta)\rho}{d_1^\tau}\lambda_1\left|\tilde{h}_{w11}\right|^2\right\}<\gamma_{\sf th}\right\}
\end{align}
After some simple algebraic manipulations, the outage probability can be rewritten as
\begin{align}\label{proof:imCSI AF high outage}
P_{S}=1&-\underbrace{P\left\{ \left|\tilde{h}_{w21}\right|^2< \frac{d_1^\tau d_2^\tau \gamma_{\sf th}}{\eta\theta\rho\sigma_1\left(\sum\limits_{i=2}^N\lambda_i\left|\tilde{h}_{w1i}\right|^2+\lambda_1\left|\tilde{h}_{w11}\right|^2\right)} \cup \left|\tilde{h}_{w11}\right|^2<\frac{\gamma_{\sf th}d_1^\tau}{(1-\theta)\rho\lambda_1}\right\}}_{P_2}\nonumber\\
&-\underbrace{P\left\{\left\|{\bf{h}}_1\right\|^2\geqslant \frac{d_1^\tau d_2^\tau \gamma_{\sf th}}{\eta\theta\rho\sigma_1\left|\tilde{h}_{w21}\right|^2}\right\}}_{P_1}+\underbrace{P\left\{\left|\tilde{h}_{w11}\right|^2<\frac{\gamma_{\sf th}d_1^\tau}{(1-\theta)\rho\lambda_1}\right\}}_{P_3}.
\end{align}
Noticing that $\sum_{i=2}^N\lambda_i\left|\tilde{h}_{w1i}\right|^2$ is a hypoexponential random variable, while $\left|\tilde{h}_{w11}\right|^2$ and $\left|\tilde{h}_{w21}\right|^2$ are i.i.d exponential random variables which are independent of $\sum_{i=2}^N\lambda_i\left|\tilde{h}_{w1i}\right|^2$, $P_1$ can be computed as
\begin{align}\label{piimcsi}
P_1=\sum_{i=1}^N\prod_{\begin{subarray}{l}j=1\\j \ne i\end{subarray}}^N\frac{\lambda_i^{N-1}}{\lambda_i-\lambda_j}\int_{0}^{\infty}e^{-\left(x+\frac{d_1^\tau d_2^\tau \gamma_{\sf th}}{\eta\theta\rho\sigma_1\lambda_i x}\right)}dx.
\end{align}
With the help \cite[Eq. (8.432.7)]{Tables} and \cite[Eq. (8.446)]{Tables}, (\ref{piimcsi}) can be evaluated as
\begin{multline}\label{p122}
P_1=\sum_{i=1}^N\prod_{\begin{subarray}{l}j=1\\j \ne i\end{subarray}}^N\frac{\lambda_i^{N-1}}{\lambda_i-\lambda_j}\left(1+\sum_{k_1=0}^\infty\frac{\left(\frac{d_1^\tau d_2^\tau \gamma_{\sf th}}{\eta\theta\rho\sigma_1\lambda_i}\right)^{k_1+1}}{k_1!(k_1+1)!}\ln{\frac{d_1^\tau d_2^\tau \gamma_{\sf th}}{\eta\theta\rho\sigma_1\lambda_i}}\right.\\
\left.-\underbrace{\sum_{k_2=0}^\infty\frac{\left(\frac{d_1^\tau d_2^\tau \gamma_{\sf th}}{\eta\theta\rho\sigma_1\lambda_i}\right)^{k_2+1}}{k_2!(k_2+1)!}\left[\psi(k_2+1)+\psi(k_2+2)\right]}_{I_0}\right).
\end{multline}

Invoking Lemma \ref{le1}, we see that the terms associated with $k_2<N-1$ in $I_0$ is zero, hence, only terms associated with $k_2\geqslant N-1$ need to be considered. Then omitting the terms containing higher order of $\frac{1}{\rho}$, $P_1$ can be approximated as
\begin{align}
P_1 \approx 1+\sum_{i=1}^N\prod_{\begin{subarray}{l}j=1\\j \ne i\end{subarray}}^N\frac{\lambda_i^{N-2}}{\lambda_i-\lambda_j}\frac{d_1^\tau d_2^\tau \gamma_{\sf th}}{\eta\theta\rho\sigma_1}\ln{\frac{d_1^\tau d_2^\tau \gamma_{\sf th}}{\eta\theta\rho\sigma_1\lambda_i}}.
\end{align}

We now look into the term $P_2$, and we have
\begin{align}\label{imCSI h2 t1}
P_2=&\sum_{i=2}^N\prod_{\begin{subarray}{l}j=2\\j \ne i\end{subarray}}^N\frac{\lambda_i^{N-3}}{\lambda_i-\lambda_j}\int_0^{\frac{\gamma_{\sf th}d_1^\tau}{(1-\theta)\rho\lambda_1}}
\int_0^{\infty}\left(1-e^{-\frac{d_1^\tau d_2^\tau \gamma_{\sf th}}{\eta\theta\rho\sigma_1(k+\lambda_1x)}}\right)e^{-\frac{k}{\lambda_i}}e^{-x}dk dx\nonumber\\
=&1-\frac{\gamma_{\sf th}d_1^\tau}{(1-\theta)\rho\lambda_1}-\sum_{i=2}^N\prod_{\begin{subarray}{l}j=2\\j \ne i\end{subarray}}^N\frac{\lambda_i^{N-3}}{\lambda_i-\lambda_j}\int_0^{\frac{\gamma_{\sf th}d_1^\tau}{(1-\theta)\rho\lambda_1}}
\int_0^{\infty}e^{-\frac{k}{\lambda_i}}e^{-\frac{d_1^\tau d_2^\tau \gamma_{\sf th}}{\eta\theta\rho\sigma_1(k+\lambda_1x)}}e^{-x}dk dx,
\end{align}
where $k=\sum_{i=2}^N\lambda_i\left|\tilde{h}_{w1i}\right|^2$ and $x=\left|\tilde{h}_{w11}\right|^2$.

Making a change of variable $t=k+\lambda_1 x$ and expanding $e^{-\frac{d_1^\tau d_2^\tau \gamma_{\sf th}}{\eta\theta\rho\sigma_1 t}}$ by the Taylor expansion, (\ref{imCSI h2 t1}) can be expressed as
\begin{align}
P_2=1-\frac{\gamma_{\sf th}d_1^\tau}{(1-\theta)\rho\lambda_1}-\underbrace{\sum_{i=2}^N\prod_{\begin{subarray}{l}j=2\\j \ne i\end{subarray}}^N\frac{\lambda_i^{N-3}}{\lambda_i-\lambda_j}\int_0^{\frac{\gamma_{\sf th}d_1^\tau}{(1-\theta)\rho\lambda_1}}e^{\left(\frac{\lambda_1}{\lambda_i}-1\right)x}
\int_{\lambda_1 x}^{\infty}\sum_{l=0}^{\infty}\left(-\frac{d_1^\tau d_2^\tau \gamma_{\sf th}}{\eta\theta\rho\sigma_1 t}\right)^{l}\frac{1}{l!}e^{\frac{-t}{\lambda_i}}dt dx}_{I}.
\end{align}

We first look at the term $l=0$, after some manipulations, we have
\begin{align}
I_{l=0}=1-\frac{\gamma_{\sf th}d_1^\tau}{(1-\theta)\rho\lambda_1}.
\end{align}

When $l = 1$, utilizing \cite[Eq. (8.211.1)]{Tables}, we have
\begin{align}\label{Il=1}
I_{l=1}=-\sum_{i=2}^N\prod_{\begin{subarray}{l}j=2\\j \ne i\end{subarray}}^N\frac{\lambda_i^{N-3}}{\lambda_i-\lambda_j}\frac{d_1^\tau d_2^\tau \gamma_{\sf th}}{\eta\theta\rho\sigma_1}\int_0^{\frac{\gamma_{\sf th}d_1^\tau}{(1-\theta)\rho\lambda_1}}e^{\left(\frac{\lambda_1}{\lambda_i}-1\right)x}{\sf Ei}\left(-\frac{\lambda_1}{\lambda_i}x\right)dx.
\end{align}

Using integration by part, with the help of \cite[Eq. (8.214.1)]{Tables}, (\ref{Il=1}) can be computed by
\begin{multline}
I_{l=1}=\sum_{i=2}^N\prod_{\begin{subarray}{l}j=2\\j \ne i\end{subarray}}^N\frac{\lambda_i^{N-2}}{\lambda_i-\lambda_j}\frac{d_1^\tau d_2^\tau \gamma_{\sf th}}{\eta\theta\rho\sigma_1}\frac{1}{\lambda_1-\lambda_i}\Bigg[(-C-\ln{\frac{\lambda_1}{\lambda_i}}-
\ln{\frac{\gamma_{\sf th}d_1^\tau}{(1-\theta)\rho\lambda_1}}-\sum_{k_1=1}^\infty\frac{\left(-\frac{\gamma_{\sf th}d_1^\tau}{(1-\theta)\rho\lambda_1}\right)^{k_1}}{k_1\times k_1!}\\+e^{\left(\frac{\lambda_1}{\lambda_i}-1\right)\frac{\gamma_{\sf th}d_1^\tau}{(1-\theta)\rho\lambda_1}}
\left.\left(C+\frac{\lambda_1}{\lambda_i}\frac{\gamma_{\sf th}d_1^\tau}{(1-\theta)\rho\lambda_1}+\sum_{k_2=1}^\infty\frac{\left(-\frac{\gamma_{\sf th}d_1^\tau}{(1-\theta)\rho\lambda_i}\right)^{k_2}}{k_2\times k_2!}\right)\right].
\end{multline}
Now expanding $e^{\frac{\lambda_1-\lambda_i}{\lambda_i}\frac{\gamma_{\sf th}d_1^\tau}{(1-\theta)\rho-\lambda_i}}$ by the Taylor expansion and omitting the higher order terms, we can obtain
\begin{align}
I_{l=1}& \approx \sum_{i=2}^N\prod_{\begin{subarray}{l}j=2\\j \ne i\end{subarray}}^N\frac{\lambda_i^{N-2}}{\lambda_i-\lambda_j}\frac{d_1^\tau d_2^\tau \gamma_{\sf th}}{\eta\theta\rho\sigma_1}\frac{1}{\lambda_1-\lambda_i}\left(-C-\ln{\frac{\gamma_{\sf th}d_1^\tau}{(1-\theta)\rho\lambda_1}}+C+\ln{\frac{\gamma_{\sf th}d_1^\tau}{(1-\theta)\rho\lambda_i}}-\ln{\frac{\lambda_1}{\lambda_i}}\right)\nonumber\\
&=0.
\end{align}
Similarly, for the terms associated with $l \geqslant 1$, it can be shown that the minimum order is $\frac{1}{\rho^2}$. Then pulling everything together, we conclude that the minimum order of $P_2$ is $\frac{1}{\rho^2}$.

Finally, we consider the term $P_3$, and we have
\begin{align}
P_3 \approx \frac{\gamma_{\sf th} d_1^\tau}{(1-\theta)\rho\lambda_i}.
\end{align}
To this end, combining $P_1$, $P_2$ and $P_3$ together yields the desired result.

\section{Proof of Proposition \ref{proposition impact of correlaion in statistical CSI}}\label{appendix:proposition impact of correlaion in statistical CSI}
It is easy to show that the end-to-end SNR given in (\ref{inscsi snr}) is a monotonic increasing function with respect to $\sigma_1$, hence, for two vectors ${\bm{\sigma}}_1$ and ${\bm{\sigma}}_2$ satisfying ${\bm{\sigma}}_1\succ{\bm{\sigma}}_2$, we have
\begin{align}
C_S({\bm{\sigma}}_1)\geqslant C_S({\bm{\sigma}}_2),
\end{align}
which completes the first half of the proof. We now turn to the impact of receive antenna correlation on the ergodic capacity.

Let ${\bm{\lambda}}$ denotes the vector containing the eigenvalues of ${\bf{R}}_r$, i.e., ${\bm{\lambda}}=\left[\lambda_{1},\lambda_{2},\cdots,\lambda_{N}\right]$, we define $\bar{\bm{\lambda}}=\left[\lambda_{2},\lambda_{3},\cdots,\lambda_{N}\right]$. According to the end-to-end SNR given in (\ref{inscsi snr}), we can express the ergodic capacity as
\begin{align}\label{CS im}
C_{s} &= \frac{1}{2}E\left[\log_2\left(1+\gamma^{\sf}_{S}\right)\right]\\
&= \frac{1}{2}E\left[\log_2\left(1+\frac{\frac{\eta\theta(1-\theta)\rho^2}{d_1^{2\tau}d_2^\tau}\left(\lambda_1\left|\tilde{h}_{w11}\right|^2+\sum_{i=2}^N\lambda_i\left|\tilde{h}_{w1i}\right|^2\right)\lambda_1
\left|\tilde{h}_{w11}\right|^2\sigma_1\left|\tilde{h}_{w21}\right|^2}{\frac{\eta\theta\rho}{d_1^\tau d_2^\tau}\left(\lambda_1\left|\tilde{h}_{w11}\right|^2+\sum_{i=2}^N\lambda_i\left|\tilde{h}_{w1i}\right|^2\right)\sigma_1
\left|\tilde{h}_{w21}\right|^2+\frac{(1-\theta)\rho}{d_1^\tau}\lambda_1\left|\tilde{h}_{w11}\right|^2+1}\right)\right]\nonumber\\
&=C_{s}\left(\bar{\bm{\lambda}},\lambda_1\right).
\end{align}
For two eigenvalue vectors ${\bm{\lambda}}_1=\left[\lambda_{11},\lambda_{12},\cdots,\lambda_{1N}\right]$ and ${\bm{\lambda}}_2=\left[\lambda_{21},\lambda_{22},\cdots,\lambda_{2N}\right]$ satisfying ${\bm{\lambda}}_1 \succ {\bm{\lambda}}_2$, we construct the following three vectors $\bar{\bm{\lambda}}_1=\left[\lambda_{12},\lambda_{13},\cdots,\lambda_{1N}\right]$, $\bar{\bm{\lambda}}_2=\left[\lambda_{22},\lambda_{23},\cdots,\lambda_{2N}\right]$ and $\bar{\bm{\lambda}}_m=\left[\lambda_{12}+\lambda_{11}-\lambda_{21},\lambda_{13},\cdots,\lambda_{1N}\right]$. According to the definition \ref{definition:majorize}, we have $\lambda_{11}\geqslant\lambda_{21}$ and $\bar{\bm{\lambda}}_m \succ \bar{\bm{\lambda}}_2$.


It is easy to prove that the second order derivative of $\log_2\left(1+\gamma^{\sf}_{S}\right)$ in (\ref{CS im}) with respect to $\sum_{i=2}^N\lambda_i\left|\tilde{h}_{w1i}\right|^2$ is small than zero, which indicates $\log_2\left(1+\gamma^{\sf}_{S}\right)$ is a concave function with respect to $\sum_{i=2}^N\lambda_i\left|\tilde{h}_{w1i}\right|^2$. Then, invoking Lemma \ref{lemma 2}, we have
 \begin{align}
 C_S\left(\bar{\bm{\lambda}}_2,\lambda_{11}\right) \geqslant C_S\left(\bar{\bm{\lambda}}_m,\lambda_{11}\right).
 \end{align}

Noticing that the first element of vector $\bar{\bm{\lambda}}_m$ is larger than the first element of vector $\bar{\bm{\lambda}}_1$, i.e.,
\begin{align}
\lambda_{12}+\lambda_{11}-\lambda_{21}>\lambda_{21},
\end{align}
and the fact that $\log_2\left(1+\gamma^{\sf}_{S}\right)$ is a monotonically increasing function with respect to $\lambda_i$, we have
\begin{align}
C_S\left(\bar{\bm{\lambda}}_m,\lambda_{11}\right)\geqslant C_S\left(\bar{\bm{\lambda}}_1,\lambda_{11}\right).
\end{align}
Hence, we have
 \begin{align}
 C_S\left(\bar{\bm{\lambda}}_2,\lambda_{11}\right)\geqslant C_S\left(\bar{\bm{\lambda}}_1,\lambda_{11}\right).
  \end{align}
When $\lambda_{11}=\lambda_{21}$, we have
 \begin{align}
 C_S\left({\bm{\lambda}}_2\right)=C_S\left(\bar{\bm{\lambda}}_2,\lambda_{11}\right)\geqslant C_S\left(\bar{\bm{\lambda}}_1,\lambda_{11}\right)= C_S\left({\bm{\lambda}}_1\right),
  \end{align}
which completes the proof.

\nocite{*}
\bibliographystyle{IEEE}
\begin{footnotesize}

\end{footnotesize}
\end{document}